\documentclass[showpacs,preprintnumbers,amsmath,amssymb, prd]{revtex4}
\usepackage{graphicx}% Include figure files
\usepackage{amsmath,amssymb,graphics,epsfig}
\usepackage{dcolumn}% Align table columns on decimal point
\usepackage{epsf}
\usepackage{epstopdf}
\usepackage {amssymb}
\newcommand{\nc}{\newcommand}
\nc{\ba}{\begin{eqnarray}}
\nc{\ea}{\end{eqnarray}}
\newcommand\be{\begin{equation}}
\newcommand\ee{\end{equation}}

\newcommand{\calH}{{\cal H}}
\newcommand{\calR}{{\cal R}}
\newcommand\mPl{{M_{\rm Pl}}}

\newcommand{\bmx}{\mathbf{x}}
\newcommand{\bmk}{\mathbf{k}}

\begin{document}

\title{Multiple Inflationary Stages with Varying Equation of State}

\author{Mohammad Hossein Namjoo$^{1, 2}$}
\email{mh.namjoo-AT-mail.ipm.ir}
\author{Hassan Firouzjahi$^{3}$}
\email{firouz-AT-mail.ipm.ir}
\author{Misao Sasaki$^{2}$}
\email{misao-AT-yukawa.kyoto-u.ac.jp}
\affiliation{$^1$School of Physics, Institute for Research in 
Fundamental Sciences (IPM),
P. O. Box 19395-5531,
Tehran, Iran}
\affiliation{$^2$Yukawa Institute for theoretical Physics,
 Kyoto University, Kyoto 606-8502, Japan}
\affiliation{$^3$School of Astronomy, Institute for Research in 
Fundamental Sciences (IPM),
P. O. Box 19395-5531,
Tehran, Iran}

\date{\today}

\begin{abstract}
\vspace{0.3cm}
We consider a model of inflation consisting a single fluid with a 
time-dependent equation of state.  In this phenomenological picture, 
two periods of inflation are separated by an intermediate non-inflationary 
stage which can be either a radiation dominated, matter dominated or kinetic 
energy dominated universe, respectively, with the equation of state 
$w=1/3$, $0$ or $1$. We consider the toy model  in which the change in 
$w$ happens instantaneously. Depending on whether the mode of interest 
leaves the horizon before or after or between the phase transitions,
the curvature power spectrum can have non-trivial sinusoidal modulations.  
This can have interesting observational 
implications for CMB anisotropies and for primordial black-hole formation.

\vspace{0.3cm}

\end{abstract}

\preprint{YITP-12-56, IPM/A-2012/012}

\maketitle

\section{Introduction}
\label{sec:introduction}

Inflation has emerged as the leading paradigm for early universe
 cosmology and structure formation. Basics predictions of inflation 
are in good agreement with cosmological observations. Namely, 
simplest models of inflation predict almost scale invariant, 
almost Gaussian and almost adiabatic 
fluctuations on cosmic microwave background (CMB) which are accurately
 measured in recent cosmological observations~\cite{Komatsu:2010fb}.
Nonetheless, it is interesting to consider more elaborate models of 
inflation which can predict observable  deviations from these simple 
predictions. In particular, models of inflation with local features 
may be interesting. Observationally, these models are employed to 
address the glitches in the CMB angular power spectrum on scales 
$\ell \sim 20-40$. Theoretically, one can construct different scenarios
 which can contain local features
\cite{Starobinsky:1992ts, Mukhanov:1991rp, Leach:2001zf, Adams:2001vc, Kaloper:2003nv, Gong:2005jr, Joy:2007na, Chen:2006xjb, Chen:2008wn, Chen:2011zf, Chen:2011tu, Pi:2012gf, Hotchkiss:2009pj, Hazra:2010ve, Jain:2008dw,  Arroja:2011yu, Adshead:2011jq, Abolhasani:2012px, Arroja:2012ae, Ackerman:2010he}.  
Models with local features may  originate from high energy physics, 
particle creations, field annihilations, change in sound speed 
or time variations of the
Newton constant during inflation~\cite{Romano:2008rr, Battefeld:2010rf, Firouzjahi:2010ga, Battefeld:2010vr, Barnaby:2009dd, Barnaby:2010ke, Biswas:2010si, Silverstein:2008sg, Flauger:2009ab, Flauger:2010ja, Bean:2008na, Nakashima:2010sa, Miranda:2012rm,Emery:2012sm, 
Abolhasani:2012tz}.
Many of these models are based on multi-field or multi-fluid scenarios.
As a consequence, there are always iso-curvature perturbations which 
may be constrained from CMB observations.

In this work we consider a phenomenological model with multiple 
inflationary stages. Different stages of inflation are separated by 
an intermediate non-inflationary period. In our model, these 
multiple inflationary stages are realized by changes in the equation of 
state $w$ for a single fluid. During inflation $w\simeq -1$ while in 
the intermediate non-inflationary stage we have $1+3w >0$. 
Particular interests are the cases in which the intermediate 
non-inflationary stage has the equation of state   $w = w_2=  1/3$, $0$
or $1$, corresponding respectively to  a radiation, matter or kinetic energy 
dominated universe. Having this said, we should emphasis that this is
 a phenomenological study and a dynamical mechanism causing the jump
 in $w$ has yet to be constructed. We shall briefly present a simple 
scalar field model which can provide a simple dynamical mechanism 
for changing $w$. Idea similar to this line of thought was studied in \cite{Allahverdi:2007ts}
in the context of MSSM inflation

The rest of the paper is organized as follows. In section \ref{back}
 we present our setup and background equations. 
In section \ref{perturbations} we present the general perturbation 
equations with appropriate matching conditions. The resulting 
transfer function of the outgoing perturbations  for arbitrary 
$w_2\neq 0$ is given in Section \ref{transfer}. 
The special case of $w_2 =0$ is considered in section \ref{matter}. 
The conclusion and discussions are given in section \ref{conclusion}
 and  some technical issues are relegated to appendices.

%%%%%%%%%%%%%%%%%%%%%%%%%%%%%%%%%%%%%%%%%%%%%%%%%%

\section{The background}
\label{back}

Here we present the background evolution of a universe filled with
a single perfect fluid with an arbitrary but constant $w=P/\rho$
where $\rho$ and $P$ are the energy density and pressure, respectively. 
The background space-time is assumed to be a flat FLRW universe,
\ba
ds^2 =  -dt^2+ a(t)^2 d \bmx^2=a^2(\eta)(-d\eta^2+d\bmx^2) \,,
\ea
where $\eta$ defined by $d\eta = dt/a(t)$ is the conformal time.

To be specific we have the following picture in mind. 
We have three distinct stages of an expanding background in which 
two inflationary periods are separated by an intermediate  
non-inflationary stage. 
The first inflationary stage continues till $\eta = \eta_{12}$ 
and during this period the fluid driving inflation has a constant 
equation of state $w= w_1$. In order to support inflation 
we require $1+ 3 w_1 <0$.
We assume that at $\eta =\eta_{12}$ the equation of state changes 
sharply from $w_1$ to $w_2$ such that  $1+ 3w_2 >0$ and 
the first stage of inflation is terminated.
As specific examples we shall consider the important cases
of $w_2 =1/3$, $0$ and $1$, corresponding respectively to 
radiation, matter and kinetic energy dominated universes. 
The third expanding stage starts at $\eta =\eta_{23}$ 
when $w$ goes a second abrupt change from $w_2$ to $w_3$. 
In order to support the final stage of inflation we assume 
that $1+ 3w_3 <0$. The time when the inflation ends is set to be
$\eta= \eta_e=0$ followed by a (p)reheating era. 
In summary, we have two inflationary stages with equations
of state $w=w_1$ and $w_3$ separated by an intermediate 
non-inflationary stage with $w= w_2$.

One may wonder how dynamically these jumps in $w$ can be realized 
in a consistent way. In  Discussions Section 
 we present a simple 
scalar field model which can mimc this behavior. 
However, in this and the following sections, we shall proceed 
phenomenologically assuming that there exists a dynamical mechanism
which can cause these changes in $w$. For our analytical analysis 
we proceed with the arbitrary sharp changes in $w$. 
We note that physically it is expected that the process in which $w$ 
undergoes large changes will take some finite lapse of time.
Therefore we also consider numerically the case when there is a short 
but finite duration of the phase transitions. 
We shall also compare our analytical results with sudden changes 
in $w$ to those obtained numerically in which the change in $w$ takes
a finite lapse of time.

With this picture in mind, now we present the background equations. 
Using the energy conservation equation and denoting the initial 
conditions with subscript $0$, the evolution of energy density  is given by
\ba
\rho= \rho_0 \left(\dfrac{a}{a_0} \right)^{-3(1+w)}\,.
\ea 
Here and below the subscript $0$ collectively denotes the time of 
phase transitions, so it corresponds to either $\eta_{12}$ or 
$\eta_{23}$ depending on which period is studied 
(see Eq.~\eqref{H} below for further details).
The Friedmann equation for a flat universe is 
\ba
3 \mPl^2 \calH^2= a^2 \rho\,,  
\ea
where ${\cal H}={a'}/{a} $ is the conformal Hubble parameter and 
the prime denotes the derivative with respect to conformal time $\eta$.
The Friedmann equation for $w \neq -\dfrac{1}{3}$ can be integrated to 
\begin{eqnarray}
\label{a-eq}
a(\eta) = a_0\Bigl(\frac{{\cal H}_0}{\beta}(\eta-\eta_0)+1\Bigr)^\beta\,,
\end{eqnarray}
where $a_0 \equiv a(\eta_0)$ and
\begin{eqnarray}
\label{beta}
\quad \beta=\frac{2}{3w+1}\,.
\end{eqnarray}
Taking the conformal time derivative of Eq.~(\ref{a-eq}),
one obtains
\ba
\calH=\frac{\calH_0}
{1+\displaystyle\frac{\calH_0}{\beta}\,( \eta- \eta_0 )}\,.
\ea
It is easy to see that both the scale factor $a(\eta)$ and
the conformal Hubble parameter ${\cal H}$ must be continuous at the time 
of phase transition $\eta=\eta_0$ when $w$ undergoes a sudden change. 

We label quantities at the three stages with 1, 2 and 3. 
For example ${\calH}_1(\eta)$ is the conformal Hubble parameter 
during the first stage whereas ${\cal H}_{12}$ is the value of 
${\cal H}(\eta)$  at the time of first phase transition $\eta=\eta_{12}$, 
${\cal H}_{12}= {\cal H}(\eta_{12})$, and so on. As a result one has 
%%%%%%%%%%%%%%%%%%%%%%%%%%%%%%%%%%%% 
\ba 
\label{H}
{\cal H}(\eta)= \begin{cases}
\dfrac{{\cal H}_{12}}{1+{\cal H}_{12} (\eta-\eta_{12})/\beta_1} 
\qquad \mathrm{for} \quad \eta < \eta_{12}\,,
\\
\\
\dfrac{{\cal H}_{12}}{1+{\cal H}_{12}(\eta-\eta_{12})/\beta_2} 
\qquad  \mathrm{for}  \quad \eta_{12} < \eta < \eta_{23}\,,
\\
\\
\dfrac{{\cal H}_{23}}{1+{\cal H}_{23} (\eta-\eta_{23})/\beta_3}
\qquad \mathrm{for}  \quad \eta > \eta_{23}\,,
\end{cases}
\ea 
in which $\beta_i$ are defined as in Eq. (\ref{beta}) with $w$ 
replaced by $w_i$ of each stage.
Here we assume that the first stage of inflation starts at $\eta \to -\infty$
and the second stage of inflation ends at $\eta_e= 0$. 
Note that the Hubble parameters at two phase transitions are related by
\ba
 {\cal H}_{23}
=\dfrac{{\cal H}_{12}}{1+{\cal H}_{12} (\eta_{23}-\eta_{12})/\beta_2}\,,
\ea
while the Hubble parameter at the end of inflation is given by
\ba
 {\cal H}_{e}
=\dfrac{{\cal H}_{23}}{1-{\cal H}_{23}\, \eta_{23}/\beta_3}\,.
\ea
Note that, as long as the intermediate non-inflationary stage
 corresponds to a universe dominated by an ordinary matter $(w>0)$ one has
\ba
{\cal H}_{12} > {\cal H}_{23}\,, \qquad   
{\cal H}_{e} > {\cal H}_{23}   \, .
\ea

Alternatively, it may be useful to work with the number of 
$e$-folds as the clock $d n= H dt = {\cal H} d \eta$. 
With the scale factor given by Eq.~(\ref{a-eq}) one obtains
\ba
n(\eta) = n_0 + \beta \ln \left( 1+ \beta^{-1} {\cal H} (\eta - \eta_0) 
\right)\,.
\ea
In particular, the number of $e$-folds of the second non-inflationary
stage $\Delta N_2 = n_{23}- n_{12}$, where $n_{23}=n(\eta_{23})$
and $n_{12}=n(\eta_{12})$, and that of the third inflationary stage
$\Delta N_3 = n_{e}- n_{23}$, where $n_{e}=n(\eta_{e})=n(0)$,
we obtain
\ba
\frac{\calH_{12}}{\calH_{23}}  
= 1+ \beta_2^{-1} {\calH}_{12} (\eta_{23} - \eta_{12})
 = e^{ \Delta N_2/\beta_2} \,,
\quad
\frac{\calH_{23}}{\calH_{e}}  
= 1- \beta_3^{-1} {\calH}_{23} \eta_{23}
 = e^{\Delta N_3/\beta_3} \,.
\ea

%%%%%%%%%%%%%%%%%%%%%%%%%%%%%%%%%%%%%%%%%%%%%%%%%%
\section{The perturbations}
\label{perturbations}

In this section we study the perturbation equations in details. 
We study the behaviors of the comoving curvature perturbations ${\cal R}$ 
or the Bardeen potential $\Phi$ which are gauge invariant.
Some technical details are described in Appendix~\ref{perts}.
For a review see, e.g. \cite{Bassett:2005xm}.

For a universe filled with a single fluid with the known equation 
of state parameter $w$ and sound speed
$c_s$ one obtains the following equation for the Fourier
space mode function of the comoving curvature perturbation:
\ba
\label{R-eq}
{\cal R}_\bmk'' + \dfrac{(z^2)'}{z^2} {\cal R}'_\bmk 
+c_s^2 k^2 {\cal R}_\bmk=0  \, ,
\ea 
where
\ba
z \equiv a(\eta) \mPl \sqrt{3(1+w)}/c_s\,.
\ea 
Note that $c_s$ is defined as $\delta P_c = c_s^2 \delta \rho_c$ 
in which the subscript c indicates that the corresponding quantities 
are measured on the comoving hypersurface (on which the fluid 4-velocity
coincides with the unit normal to the hypersurface).

One can easily solve Eq.~(\ref{R-eq}) in each phase with constant 
values of $w$ and $c_s$. At the time of transition,
we need two matching conditions in order to match the 
outgoing solutions to the incoming solutions \cite{Deruelle:1995kd}.

The first matching condition is the continuity of the curvature perturbation 
itself,
\ba
\label{R-bc1}
[{\cal R}_\bmk]_-^+ =0   \, ,
\ea
where  $[X]_-^+$ denotes the difference in the value of quantity $X$ 
after and before the transition: $[X]_-^+ = X(\eta_+) - X(\eta_-)$.
Geometrically, the continuity of ${\cal R}$ can be interpreted as the
 continuity of the extrinsic and intrinsic curvature at the 
three-dimensional spatial hyper-surfaces located at 
$\eta=\eta_{12}$ and $\eta=\eta_{23}$, to be consistent with
the Bianchi indentity.

We also need another matching condition for the time derivative
 of ${\cal R}$. Note that the Eq.~(\ref{R-eq}) can be rewritten by
\ba
\dfrac{d}{d\eta} \left(\frac{a^2}{c_s^2} (1+w) {\cal R}'_\bmk \right)
+a^2 (1+w)  k^2 {\cal R}_\bmk=0  \, .
\ea
By integrating the above equation in a small range around the phase
 transition, the last term vanishes and one obtains the second 
matching condition by
\ba
\label{matchR'}
\left[ \dfrac{1+w}{c_s^2} {\cal R}_\bmk'\right]_\pm =0 \, .
\ea
Alternatively, one can obtain the matching condition (\ref{matchR'}) 
in a different way. From the continuity of the extrinsic and 
intrinsic three-dimensional hyper-surface at the time of phase 
transition, we also conclude that the curvature perturbation on the 
shear-free hypersurface (Newton gauge) $\Phi$ is continuous across 
the transition surface,
\ba
\label{Phi-bc}
\left[ \Phi \right]_-^+ =0 \, .
\ea
Then by noting the relations between $\Phi$ and ${\cal R}$ (see Appendix \ref{perts}
for details )
\ba
\label{RPhi}
\calR = \dfrac{2 \Phi' + (5+3w) \calH \Phi }{3 \calH (1+w)}\,,
\ea
and
\ba
\label{PhiR}
\Phi  = -  \dfrac{3 (1+w) \calH}{2 c_s^2 k^2} \calR' \,,
\ea
we see from Eq.~(\ref{PhiR}) that the matching condition~(\ref{matchR'})
implies the continuity of $\Phi$.

Now we solve the equation of motion for ${\cal R}_\bmk$.
For a constant $w$ and $c_s$, we have $z'/z=a'/a=\calH$.
Hence Eq.~(\ref{R-eq}) simplifies to
\begin{eqnarray}
\left[\frac{d^2}{dx^2}+\frac{2\beta}{x}\frac{d}{dx}+1\right]{\cal R}_\bmk=0  \, ,
\end{eqnarray}
where 
\begin{eqnarray}
\label{xdef}
x\equiv\frac{\beta}{|\beta|}c_sk(\eta-\eta_0+\beta\calH_0^{-1})\,.
\end{eqnarray}
The solution is given by
\ba
\label{R-sol}
{\cal R}_\bmk
= x^{\nu} \left[ C_1  H^{(1)}_\nu\left(x\right) 
+D_1 H^{(2)}_\nu\left(x\right) \right] \,;
\quad \nu\equiv \frac{1}{2}-\beta=\frac{3(w-1)}{2(3w+1)}  \,,
\ea
where $C_1$ and $D_1$ are constant of integrations,
and $H^{(1)}_\nu(x)$ and $H^{(2)}_\nu(x)$ are the Hankel functions of 
the first and second kinds, respectively. 
Note that during inflation $\beta <0$ (for slow-roll inflation 
$\beta \simeq -1$) and the above general definition of 
$x$ yields 
\begin{eqnarray}
x=x_1(\eta)\equiv - c_{s1} k (\eta-\eta_{12}+\beta_1\calH_{12}^{-1})\,,
\label{x1def}
\end{eqnarray}
during the first period of inflation.

Our goal is to find the curvature perturbation ${\cal R}_\bmk$ at
 the end of inflation $\eta=0$.
The power spectrum ${\cal P}_{{\calR}}$ is defined by
\ba
\label{R-powr}
\langle  {\cal R}_{ \bmk} {\cal R}_{ \bmk'}\rangle 
\equiv (2\pi)^{3} P_{{\cal R}}(k)~\delta^3(\bmk+\bmk')\,,
\quad
{\cal P}_{\cal R}\equiv \frac{k^{3}}{2 \pi^{2}}P_{\cal R}(k)\,;
\quad P_{\calR}(k)=|\calR_\bmk|^2\,,
\ea
where $\calR_\bmk$ is the normalized positive frequency mode function.
At sufficiently early times, the solution should approach 
the Minkowski positive frequency mode function. That is,
for $\eta \to -\infty$,
\ba
{\cal R}_\bmk \to  \dfrac{e^{-i c_s k \eta}}{z(\eta) \sqrt{2 c_s k}}
\quad 
\mathrm{for}  \ \eta\to -\infty\,.
\ea
Imposing this initial condition on the solution (\ref{R-sol}) and 
using the asymptotic form of the Hankel function given by 
Eq.~(\ref{large-Hankel}), we find at the first stage of inflation,
$D_1=0$ and
\ba
\label{R1}
{\cal R}_\bmk(\eta)=C_1\,
x_1(\eta)^{\nu_1} H^{(1)}_{\nu_1}(x_1(\eta))  \, ;
\quad \eta<\eta_{12}\,,
\ea
where
\ba 
C_1 \equiv  \dfrac{-1}{2 \mPl \, a(\eta_{12})}
\left( \dfrac{ i \pi c_{s1} e^{i \pi \nu_1}}{3 k (1+w_1)} \right)^{1/2}
 x_{1}(\eta_{12})^{1/2-\nu_1} \, .
\ea

Considering the slow-roll limit in which  $w_1 = -1+2\epsilon_1/3$ and 
$\nu_1\simeq 3/2$ the above equation results in the following power spectrum 
at $\eta=\eta_{12}$
for the modes which leave the horizon during the first stage of inflation:
\ba
{\cal P}_{{\calR}}(\eta_{12}) 
\simeq \dfrac{H_{12}^2}{8 \pi^2 \,\mPl^2  c_{s1} \, \epsilon_1}\,,
\ea 
where the slow-roll parameter $\epsilon$ is defined by 
\begin{eqnarray}
\epsilon \equiv - \frac{\dot H}{H^2}=\frac{3}{2}(1+w)\,.
\label{epsilondef}
\end{eqnarray}
For those modes that remain superhorizon until the end of inflation,
the curvature perturbation is conserved, and we have
\begin{eqnarray}
{\cal P}_{{\calR}}(\eta_{e})= {\cal P}_{{\calR}}(\eta_{12})
\simeq \dfrac{H_{12}^2}{8 \pi^2 \,\mPl^2  c_{s1} \, \epsilon_1}\,.
\label{PR1}
\end{eqnarray}
This is the standard results for single field inflation~\cite{Garriga:1999vw}.

Applying the general solution (\ref{R-sol}) to
the second and third stages, we have 
\ba
\label{R2}
{\cal R}_\bmk(\eta) 
&=& C_2 \, x_2^{\nu_2} H^{(1)}_{\nu_2}\left(x_2(\eta) \right)
+ D_2 \, x_2^{\nu_2} H^{(2)}_{\nu_2}\left(x_2(\eta) \right) \,;
\quad \eta_{12}<\eta<\eta_{23}\,,
\\
\label{R3}
{\cal R}_\bmk(\eta) 
&=& C_3 \, x_3^{\nu_3} H^{(1)}_{\nu_3}\left(x_3(\eta) \right)
+ D_3 \, x_3^{\nu_3} H^{(2)}_{\nu_3} \left(x_3(\eta) \right) \,;
\quad \eta_{23}<\eta \,,
\ea
where $x_i(\eta)$ ($i=2$, $3$) are defined in accordance with
the general definition~(\ref{xdef}),
\begin{eqnarray}
x_2&\equiv& c_{s2}k(\eta-\eta_{12}+\beta_2\calH_{12}^{-1})\,;
\quad \nu_2=\frac{1}{2}-\beta_2=\frac{3(w_2-1)}{2(3w_2+1)}\,,
\cr
\cr
x_3&\equiv& c_{s3}k(\eta-\eta_{23}+\beta_3\calH_{23}^{-1})\,;
\quad \nu_3=\frac{1}{2}-\beta_3=\frac{3(w_3-1)}{2(3w_3+1)}\,.
\end{eqnarray}
Here we assume that $w_2=c_{s2}^2\neq 0$, so the intermediate
 non-inflationary stage is not a matter-dominated universe. 
The case when the intermediate stage is matter dominated with
$w= c_{s2}^2=0$ is considered separately in Section \ref{matter}.

As usual, we are interested in modes which are super-horizon at
the end of inflation $x_3(\eta_e) \ll 1$ where $\eta_e \rightarrow 0$.
Using the asymptotic form of the Hankel function, we obtain
\ba
 \label{R-end}
{\cal R}_\bmk(\eta \to 0) \simeq - \dfrac{i\,  2^{\nu_3}}{\pi}\,
 \Gamma(\nu_3)   (C_3-D_3)\,.
\ea
It is useful to define the transfer function $T$ for the power spectrum 
as
\ba
{\cal P}_{\cal R}(\eta=0) = T\,{\cal P}_{{\cal R}_1}(\eta=0) \,,
\ea
where ${\cal P}_{{\calR}_1}(\eta=0)$ is the power spectrum at 
the end of inflation if there were no transition
and $w=w_1$ throughout the inflationary stage,
as calculated in Eq. (\ref{PR1}).
Assuming $\nu_1 \simeq \nu_3 \simeq 3/2$ we obtain
\ba
\label{T-def}
T \simeq \dfrac{\vert C_3 - D_3 \vert^2}{\vert C_1 \vert^2}  \, .
\ea
Thus any non-trivial effect due to change in $w$ is captured by 
a non-trivial transfer function $T \neq 1$.  
The details of the calculation of the coefficients $C_2$, 
$D_2$, $C_3$ and $D_3$ are given in Appendix~\ref{transfer-app}.

%%%%%%%%%%%%%%%%%%%%%%%%%%%%%%%%%%%%%%%%%%%%%%%%%% 
\section{Transfer Function}
\label{transfer}

In this section we calculate the transfer function $T$ which encodes 
the effects of change in $w$. As mentioned above, in this section we 
assume that the intermediate non-inflationary stage is not 
matter dominated so $w_2=c_{s2}^2 \neq 0$ and our 
formulas~(\ref{exact1})-(\ref{exact4}) are valid. 
The case in which $w_2= c_{s2}^2 =0 $ is studied separately in 
Section~\ref{matter}. 
Also we provide the general formula for arbitrary $w_2 \neq 0$ and 
then consider the particular example $w_2=1/3 $ and $w_2 =1$ corresponding,
respectively, to a radiation dominated and kinetic energy dominated universe. 
  
\begin{figure}
\includegraphics[width =  3.5in ]{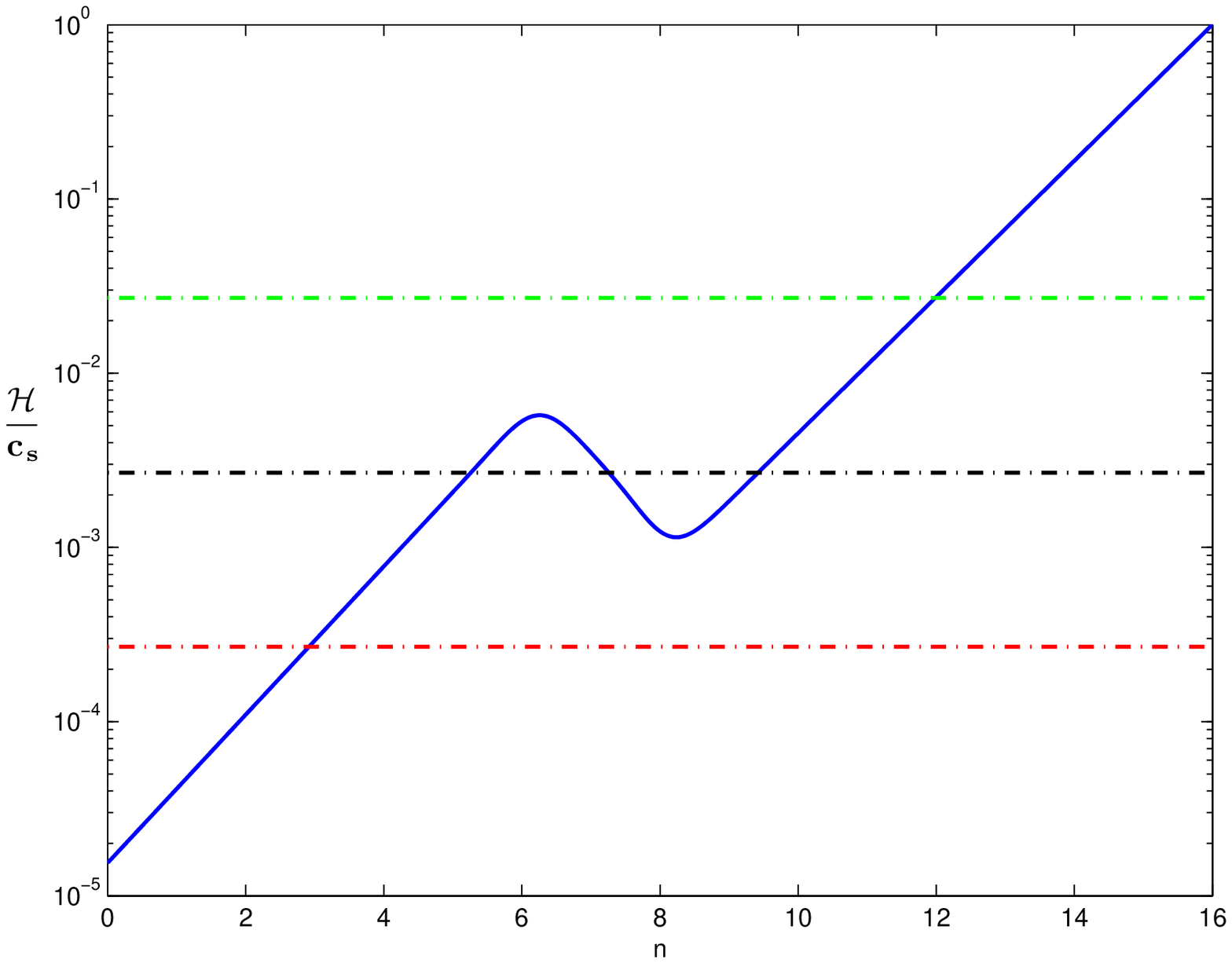}
\includegraphics[width =  3.5in ]{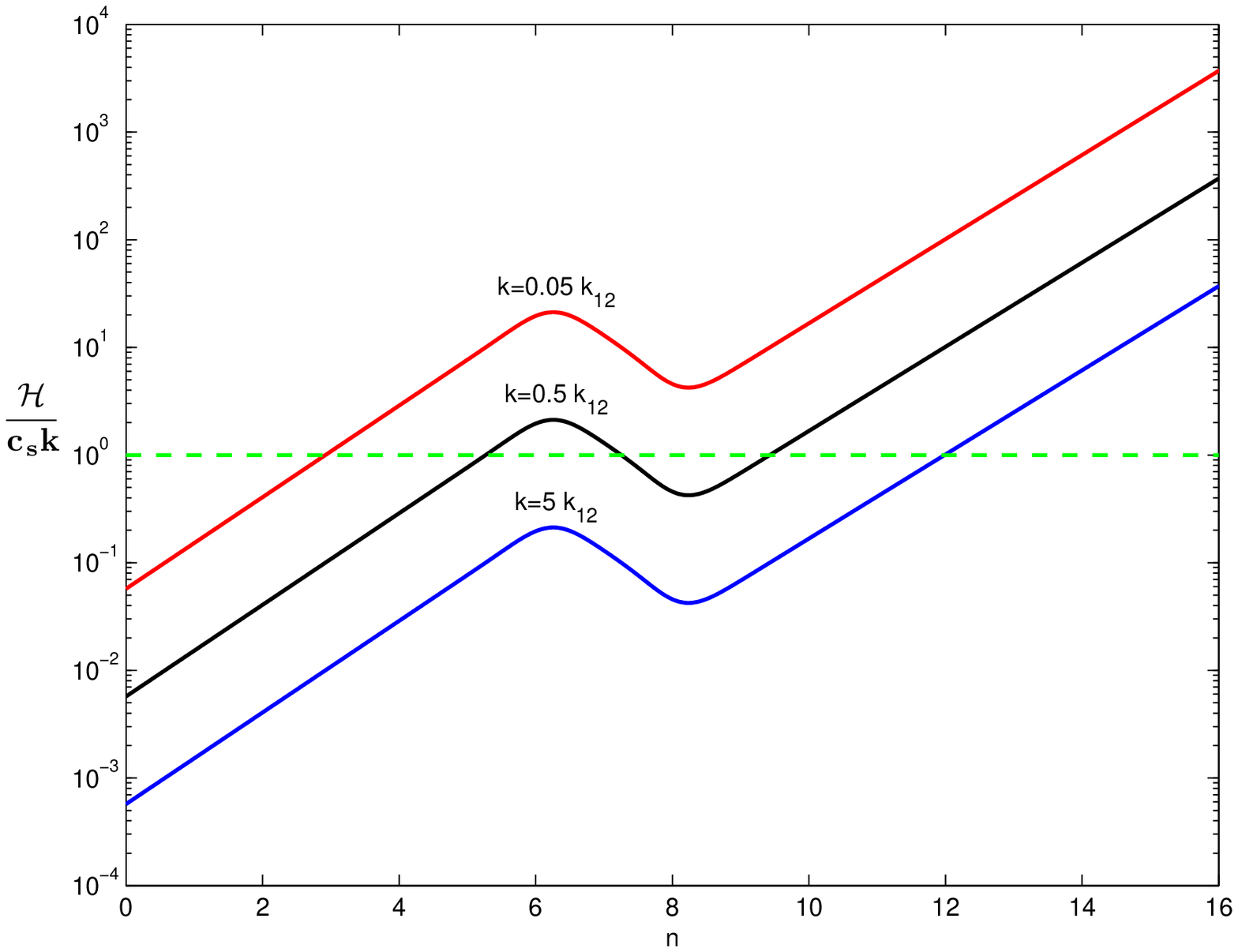}
\caption{
Here we plot ${\cal H}/c_s$ (left) and the dimensionless number 
${\cal H}/k\, c_s$ (right) as a function of the number of $e$-folds $n$.  
As usual, during  inflation the ratio ${\cal H}/c_s$ increases while
 during the second stage (non-inflationary period) it decreases.  
In the left plot, 
the upper, middle and lower horizontal lines correspond, respectively, to
 $k= 5 k_{12}$, $k=0.5k_{12}$ and $k= 0.05k_{12}$. 
As can be seen in both plots these modes enter the horizon at different times.  
\label{background}
}
\end{figure}

The behaviors of the power spectrum depends on the ratio $k {\cal H}/c_s$. 
Depending on this value different situations arise.
 To be specific let us define  
\ba
\label{k12-k13}
k_{12} \equiv \frac{-{\cal H}_{12}}{\beta_1 c_{s1}} \,,
\quad 
k_{23} \equiv \frac{-{\cal H}_{23}}{\beta_3 c_{s3}}\,.
\ea
Here $k_{12}$ represents  the mode which leaves 
the horizon at the time of the first phase transition whereas 
 $k_{23}$ is the mode which leaves the horizon at time of 
the second phase transition. Note that $k_{12} > k_{23}$
as long as the second stage is non-inflationary.
With the above definitions of the characteristic wave numbers,
there are three categories of the modes.
The behavior of ${\cal H}/c_s$ as well as that of the ratio 
of ${\cal H}/c_s$ to $k$ for three typical values of $k$ is depicted 
as a function of the number of $e$-folds in Fig.~\ref{background}.

The first category contains the modes $k > k_{12}$,
which remain sub-horizon until the second stage of inflation,
For this category we have
 $ x_1(\eta_{12})$, $x_2(\eta_{12})$, $x_2(\eta_{23})$,  
$x_3(\eta_{23})\gg 1$. 

The second category contains the modes $k_{23} < k < k_{12}$,
which leave the horizon during the first inflationary stage, 
re-enter the horizon during the second non-inflationary stage,
and finally exit the horizon during the second inflationary
stage. For this category we have
$x_1(\eta_{12})$, $x_2(\eta_{12}) \ll 1 $ and 
$x_2(\eta_{23})$, $x_3(\eta_{23}) \gg 1$.

The third category contain the modes $k< k_{23}$,
which leave the horizon during the first stage and always remain 
super-horizon until the end of inflation.
For this category we have
$x_1(\eta_{12})$, $x_2(\eta_{12})$, $x_2(\eta_{23})$, $x_3(\eta_{23}) \ll 1$.

%%%%%%%%%%%%%%%%%%%%%%%%%%%%%%%%%%%%%%%%%%%%%%%%%
%%%%%%%%%%%%%%%%%%%%%%%%%%%%%%%%%%%%%%%%%%%%%%%%

Now we study each category of the modes in turn. Let us start with the 
first category,  modes which remain subhorizon until the second phase of
 inflation, $k > k_{12}$ 
(for which 
$ x_1(\eta_{12})$, $x_2(\eta_{12})$, $x_2(\eta_{23})$, $x_3(\eta_{23})\gg 1$).
The coefficients $C_2$, $D_2$, $C_3$ and $D_3$ 
are calculated in Appendix~\ref{transfer-app}. The amplitude of the
curvature perturbation at the end of inflation is
obtained as
\ba
\label{R-cat1}
\vert {\cal R}_k(\eta=0) \vert
&\simeq&
 f_{23}\dfrac{C_1}{\pi} 2^{\nu_3-\frac{1}{2}} \Gamma(\nu_3)
\dfrac{ x_1(\eta_{12})^{\nu_1-\frac{1}{2}}}{ x_3(\eta_{23})^{\nu_3-\frac{1}{2}}} 
\left( \dfrac{ x_2(\eta_{23})}{ x_2(\eta_{12})} \right)^{\nu_2-\frac{1}{2}}  
\cr
\cr
&&\times \sqrt{1+\sin\left(2 x_3(\eta_{23}) - \pi \nu_3 \right) } 
\sin\bigl(\, x_2(\eta_{12})-x_2(\eta_{23})\bigr)\,.
\ea
Then the transfer function $T $ is calculated to be
\ba 
T^{1/2}&\simeq &
\sqrt{2} f_{23} 
\dfrac{ x_1(\eta_{12})^{\nu_1-1/2}}{ x_3(\eta_{23})^{\nu_3-\frac{1}{2}}} 
\left( \dfrac{ x_2(\eta_{23})}{ x_2(\eta_{12})} \right)^{\nu_2-\frac{1}{2}}  
 \sqrt{1+\sin\left(2 x_3(\eta_{23}\right) - \pi \nu_3 ) } 
\sin(x_2(\eta_{12})-x_2(\eta_{23}))  
\cr
\cr 
&\simeq & \sqrt{2}  f_{23}   \left(\frac{k}{k_{12}} \right)^{\nu_1-\frac{1}{2}}
 \left(\frac{k}{k_{23}} \right)^{\frac{1}{2}-\nu_3}  
\left(  \frac{\beta_1 c_{s 1}}{\beta_3 c_{s 3}} 
 \frac{k_{12}}{k_{23}}  \right)^{\nu_2-\frac{1}{2}}
\cr
\cr
&&\times \sqrt{1+ \sin \left( \frac{2k}{k_{23}} - \pi \nu_3 \right)}
 \sin \left( \frac{-\beta_2 c_{s\, 2}}{\beta_1 c_{s 1}} \frac{k}{k_{12}} 
 \left[ 1-   \frac{\beta_1 c_{s\, 1}}{\beta_3 c_{s 3}}  \frac{k_{12}}{k_{23}} 
  \right] \right)\,,
\label{T-cat1}
\ea
where we have defined 
\ba
\label{fij}
f_{ij}=  \frac{\mathrm{sgn}(1+ 3 w_i)}{\mathrm{sgn}(1+ 3w_j)}
    \dfrac{(1+w_i)}{(1+w_j)}  
\frac{c_{sj}}{c_{si}} \,,
\ea
in which $\mathrm{sgn}(x)$ is the sign function; $\mathrm{sgn}(x)=+1$ $(-1)$ 
for $x>0$ $(x<0)$.
 Note that in the continuous limit where  $w_i \rightarrow w_j$ 
we obtain the expected result that $f_{ij}\rightarrow  1$ 
corresponding to no sharp transition. Since a change in $w$ 
naturally causes a change in $c_s$ too, it is the combination $f_{ij}$ 
which controls whether or not we have a non-trivial phase transition.

Now consider the second category of the modes $k_{23} < k < k_{12}$, 
which leave the horizon at the first stage of inflation, re-enter 
during the intermediate stage and cross the horizon again during 
the second stage of inflation
(for which $x_1(\eta_{12})$, $x_2(\eta_{12}) \ll 1 $ while 
$x_2(\eta_{23})$, $x_3(\eta_{23}) \gg 1$). 
From the result given in Appendix~\ref{transfer-app}
we obtain
\ba
\label{Rsecond}
\vert {\cal R}_k (0) \vert &\simeq &
\dfrac{ \vert C_1 \vert }{\pi^2} f_{23} 2^{\nu_1-\nu_2+ \nu_3} 
\Gamma(\nu_1) \Gamma(-\nu_2+1) x_3(\eta_{23})^{-\nu_3+\frac{1}{2}}
x_2(\eta_{23})^{\nu_2-\frac{1}{2}} 
\cr
\cr
&& \times \sqrt{\left(1-\sin(2 x_2(\eta_{23}) - 3\nu_2 \pi) \right) 
\left(1+\sin(2 x_3(\eta_{23})-\pi \nu_3 )\right)}\,.
\ea
The transfer function is given by
\ba 
T^{1/2} &\simeq& 
\dfrac{ 2^{\nu_3- \nu_2}  }{\pi}f_{23} \Gamma(1-\nu_2)\, 
 x_3(\eta_{23})^{-\nu_3+\frac{1}{2}} \, x_2(\eta_{23})^{\nu_2-\frac{1}{2}}
\cr
\cr
&& \times
 \sqrt{\left(1-\sin\left(2 x_2(\eta_{23} \right) - 3\nu_2 \pi) \right) \left(1+\sin\left(2 x_3(\eta_{23}\right)-\pi \nu_3 )\right)}  
\cr
\cr
&\simeq&
\dfrac{ 2^{\nu_3- \nu_2}  }{\pi} f_{23} 
\Gamma(1-\nu_2) \left( \frac{k}{k_{23}}\right)^{-\nu_3 + \frac{1}{2}} 
\left( \frac{\beta_2 c_{s\, 2}}{\beta_3 c_{s\, 3}} \frac{k}{k_{23}}
\right)^{\nu_2 - \frac{1}{2}}
\cr
\cr
&& \sqrt{1- \sin \left(   \frac{-2\beta_2 c_{s\, 2}}{\beta_3 c_{s\, 3}}
\frac{k}{k_{23}} - 3 \nu_2 \pi \right)}
 \sqrt{1+ \sin \left( \frac{2 k}{k_{23}}- \nu_3 \pi\right)}\,.
\label{Tsecond}
\ea

Finally, consider the third category, $k< k_{23}$,
corresponding to the modes which leave the horizon during
the first stage of inflation and never re-enter the horizon 
until the end of inflation. Since the curvature perturbation
is conserved on superhorizon scales, the amplitude of these
modes are simply given by the standard result given in Eq.~(\ref{PR1}),
and the transfer function is trivial; $T\simeq1$.

\begin{figure}
\includegraphics[width =  4.5in ]{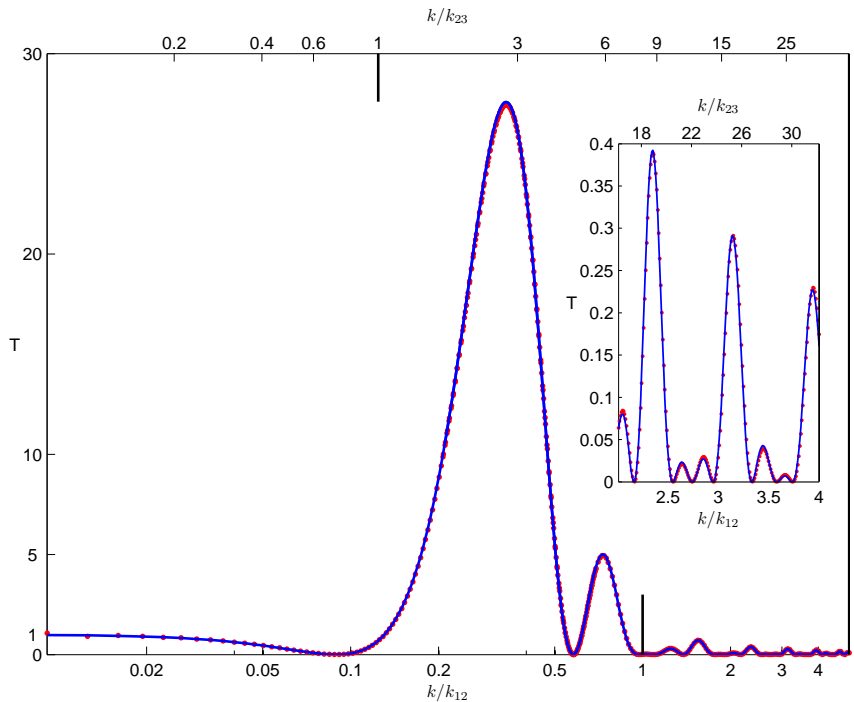}
\caption{
The transfer function $T$ is plotted for the case where $w_2=1/3$, 
i.e. the non-inflationary stage is radiation dominated.  
The numerical parameters are 
$w_1= -0.99$, $w_3= -0.93$,  $c_{s1}= c_{s\, 3}=1$, $c_{s\, 2} = \sqrt{w_2}$.
Three distinct categories are recognized as discussed below Eq.~(\ref{k12-k13}),
 corresponding to 
$k> k_{12}$,  $k_{23}<  k < k_{12}$  and $k< k_{23}$. 
The blue curve is from our analytical solution~(\ref{T-def}) while 
the red dots are from the full numerical result. 
\label{T-radiation} }
\end{figure}

The results obtained above show that neither the amplitude nor 
the scale-dependence of the power spectrum is the same as the standard case
for the modes $k>k_{23}$.
A typical example of the transfer function is shown in Fig.~\ref{T-radiation}.
The power spectrum is highly oscillatory as a function of momentum. 
This  non-trivial behavior is a result of the scattering of the initial 
wave function by the two phase transitions which lead to 
a mixing of the negative frequency modes, which are absent initially.
As a result the subhorizon modes at the second phase of inflation are 
no longer purely positive frequency. 

\begin{figure}
\includegraphics[width =  4.5in ]{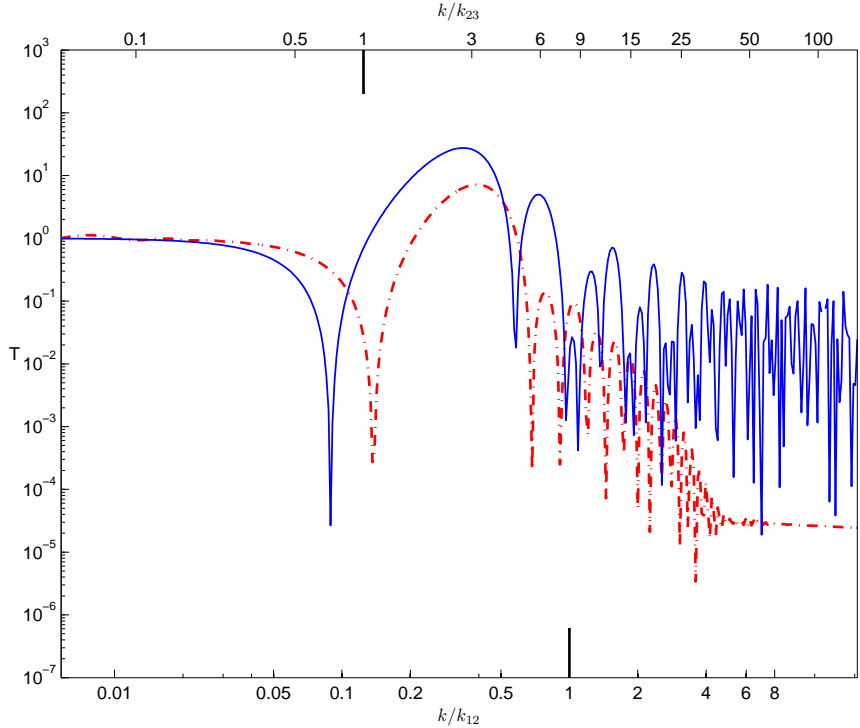}
\caption{
The transfer function $T$ (the red dashed line)  for the same situation as 
in Fig.~\ref{T-radiation} above but with a relatively mild phase
 transition in which the change in $w$ takes place in $1/3$ of an e-fold. 
As expected, for sufficiently small scales the sinusoidal modulations 
disappear. The blue solid curve is the analytic expression for the 
sharp phase transition.   
\label{T-mild}
}
\end{figure}

We note that in the limit $k \gg k_{12}$ we obtain the scaling property 
$T \propto k^{2 \nu_1 - 2 \nu_3}$ from Eq.~(\ref{T-cat1}).
The non-decaying sinusoidal modulation on top of this mild scale-dependence 
is because of the assumption that the changes in $w$ take place abruptly. 
Under this assumption the small scale modes,
no matter how deep inside the horizon they are, are all affected.
However, if we allow a finite time-scale for the change in $w$, 
say $\Delta \tau = \delta $, then the sinusoidal modulations 
on the power spectrum die out for frequencies bigger than $\delta^{-1}$
and the power spectrum reaches its almost scale-invariant value at the end 
of inflation. This behavior is seen in Fig.~\ref{T-mild}, 
which is obtained numerically for an example of smooth
changes in $w$.

%%%%%%%%%%%%%%%%%%%%%%%%%%%%%%%%%%%%%%%%%%%%%%%%%
%\vspace{0.5cm}
%%%%%%%%%%%%%%%%%%%%%%%%%%%%%%%%%%%%%%%%%%%%%%%%%%
%%%%%%%%%%%%%%%%%%%%%%%%%%%%%%%%%%%%%%%%%%%%%%%%

%%%%%%%%%%%%%%%%%%%%%%%%%%%%%%%%%%%%%%%%%%%%%%%%%
\begin{figure}
\includegraphics[width =  4.5in ]{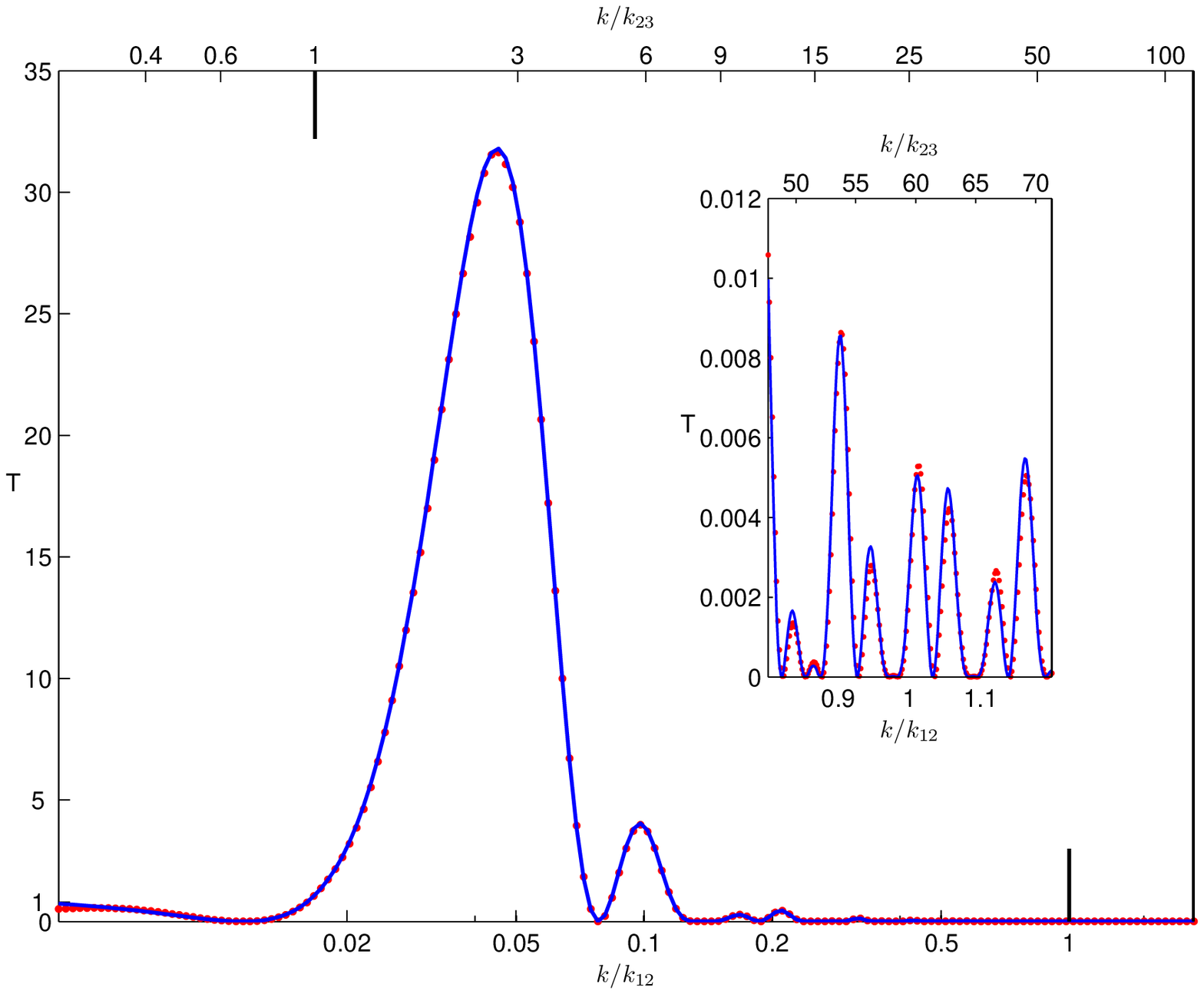}
\caption{
The transfer function $T$ is plotted for the case where $w_2=1$, i.e. 
the intermediate non-inflationary stage is a kinetic energy dominated
 universe.  The other numerical parameters are the same as in 
Fig.~\ref{T-radiation} with the same physical interpretations 
for the modes behaviors.  
\label{T-kinetic}
}
\end{figure}
%%%%%%%%%%%%%%%%%%%%%%%%%%%%%%%%%%%%%%%%%%%%%%%%

Before closing this section, let us explore the 
dependence of the power spectrum enhancement as a function of 
the duration of the intermediate non-inflationary stage, $\Delta N_2$. 
Here we propose two methods to see this behavior in which each has its
 own advantages. We especially concentrate on Eq.~\eqref{Tsecond} which
 to good approximation shows the behavior we are looking for. 
Firstly note that for a fixed value of ${\cal H}_e$  and assuming
 $w_3 \simeq -1$, the value of  $k_{23}$ is nearly determine by 
$\Delta N_2$ and as a result we would like to compare the values of 
$T$ for different $\Delta N_2$ at the mode $k= k_{23}$. 
This is the extreme limit of the validity of Eq.~\eqref{Tsecond} which 
holds for the modes in the range $k_{23} \ll k \ll k_{12} $.
Obviously, one should have $k_{23} \ll k_{12} $ if one uses the above
approximation, which is the case when $\Delta N_2$ is sufficiently large. 
 
\begin{figure}
\includegraphics[scale=.7]{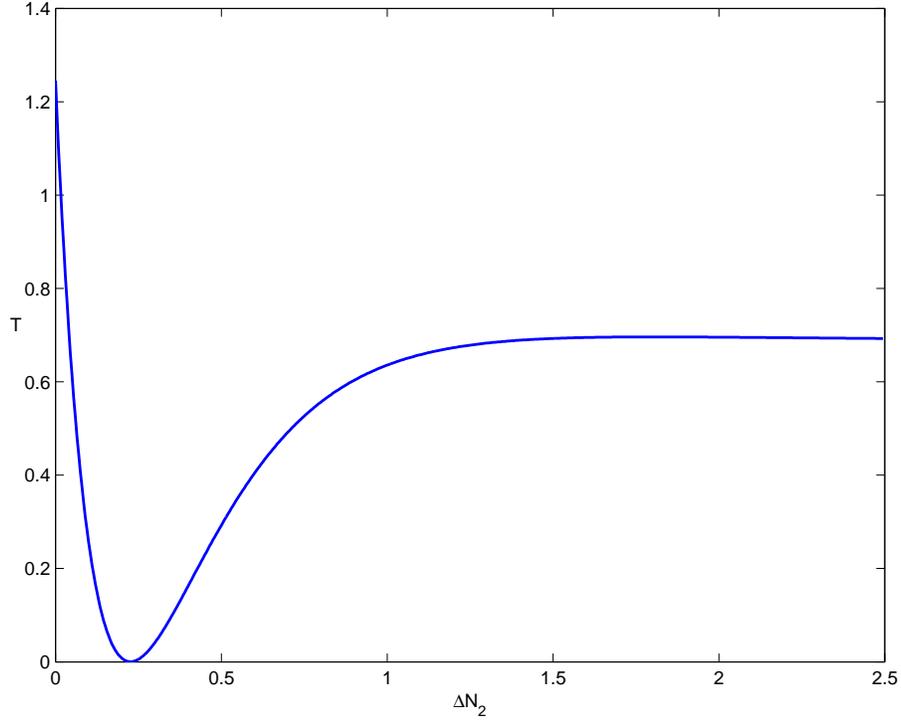} 
\caption{$T$ at the scale $k=k_{23}$ as a function of $\Delta N_2$, 
for the fixed values of $w_1$, $w_2$ and $c_{s1}$, $c_{s3}$. 
Except $\Delta N_2$, the other parameters are the same as the previous 
plots for $w_2=1/3$.}
\label{constN2}
\end{figure}

The first approach is to set the parameters $w_1$ and $w_3$ to constant
values and look for the $\Delta N_2$ dependence in the transfer 
function. An inspection of Eq.~\eqref{Tsecond} shows that for
 the scale $k=k_{23}$, there is no $\Delta N_2$ dependence on 
the amplitude of the transfer function. As a result, the transfer function
will not vary much as a function of $\Delta N_2$. This argument is
supported by Fig.~\ref{constN2}, in which the value of the
transfer function at $k=k_{23}$ is shown as a function of $\Delta N_2$.

\begin{figure}
\includegraphics[scale=.7]{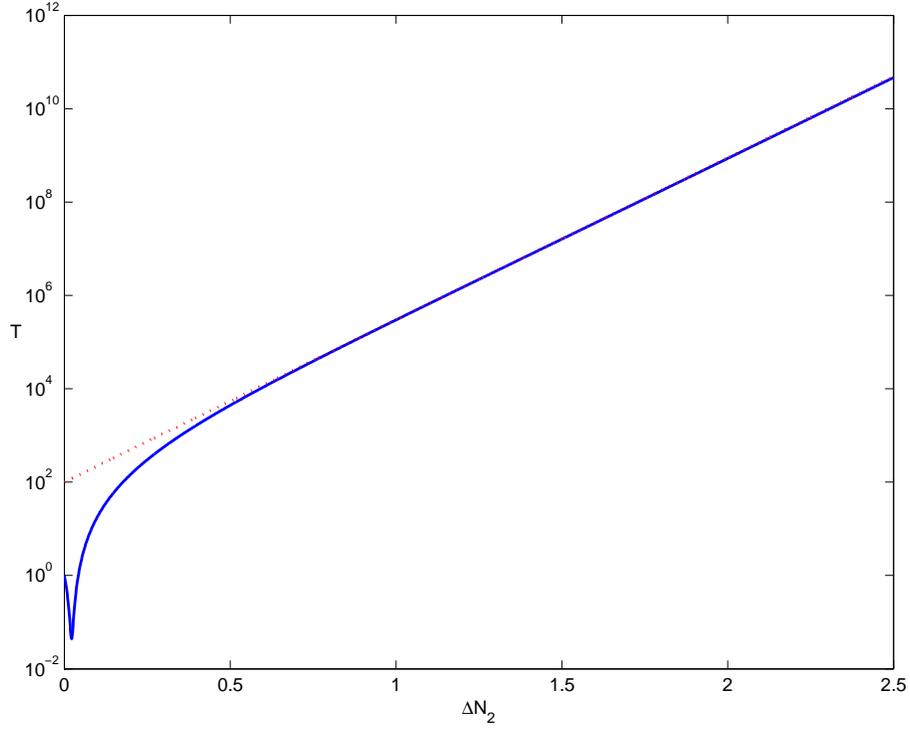} 
\caption{$\ln T$ at the scale $k=k_{23}$ as a function of $\Delta N_2$, 
for the case with equal values of  power spectrum in two stages of inflation. 
The dashed red line is a line proportional to $e^{4 N_2 (1+1/\beta_2)}$.}
\label{varN2}
\end{figure}

The second approach is to set the power spectrum of the two inflationary 
stages equal to each other and vary  $\Delta N_2$. 
In this case we impose the condition,
\ba 
 \dfrac{H_{12}^2}{\epsilon_1 c_{s1}} 
\simeq \dfrac{H_{23}^2}{\epsilon_3 c_{s3}} \,.
\ea
As a result, we should change e.g.  $\epsilon_3$ for fixed values 
of $\epsilon_1$ and $c_{si}$, when we vary $\Delta N_2$,
since the Hubble parameter changes considerably during the 
intermediate stage. In fact, using the above equality, one has
\ba 
\label{e3}
 \epsilon_3 \simeq \epsilon_1 e^{-2 \Delta N_2 (1+1/\beta_2)}
\ea
Noting that $1+w_3 =  2 \epsilon_3/3$, and $f_{23}  \propto 1/(1+w_3)$, 
one can conclude from Eq.~\eqref{Tsecond} that the transfer function 
behaves as $T \propto e^{4 \Delta N_2 (1+1/\beta_2)}$, i.e.  
the enhancement is exponentially larger for larger values of 
$\Delta N_2$, which is a very interesting phenomenon.
 This feature is also supported from the exact numerical result
shown in Fig.~\ref{varN2}.

%%%%%%%%%%%%%%%%%%%%%%%%%%%%%%%%%%%%%%%%%%%%%%%%%%
%%%%%%%%%%%%%%%%%%%%%%%%%%%%%%%%%%%%%%%%%%%%%%%%%%

%%%%%%%%%%%%%%%%%%%%%%%%%%%%%%%%%%%%%%%%%%%%%%%%%%
%%%%%%%%%%%%%%%%%%%%%%%%%%%%%%%%%%%%%%%%%%%%%%%%%%

%%%%%%%%%%%%%%%%%%%%%%%%%%%%%%%%%%%%%%%%%%%%%%%%%%
\section{matter-dominated intermediate stage} 
\label{matter}

The analysis in the previous sections are valid as long as the sound 
speed and equation of state in each stage do not vanish. However, if the 
intermediate stage is a matter-dominated universe  then the  previous 
results are not applicable since the matching conditions as well as the
 equation of motion of curvature perturbation are singular. 
In this section we obtain the power spectrum of curvature perturbation 
for this case.

In this special case, it is better to work with the curvature perturbation
on the Newtonian slice $\Phi$ (i.e., the so-called Bardeen potential),
since the equations behave properly when written in
 terms of $\Phi$. The relations between ${\cal R}$ and $\Phi$
are given in Eqs.~(\ref{RPhi}) and (\ref{PhiR}). Eliminating ${\cal R}$ 
from these formulas results in the following second order differential
 equation for $\Phi$:
\ba
\label{Phi-matter}
 \Phi'' + 3 \calH (1+c_w^2) \Phi' 
+ [c_s^2 k^2 - 3 (w - c_w^2) \calH^2 ] \Phi=0\,,
\ea
where
\ba
c_w^2 \equiv \dfrac{P'}{\rho'} = w - \dfrac{w'}{3(1+w) \calH}  \, .
\ea
It is crucial to realize that $c_w$ is not the same as $c_s$ in general. 
Note that $c_s$ is defined as $\delta P_c = c_s^2 \delta \rho_c$ 
where $\delta P_c$ and $\delta \rho_c$ are the adiabatic pressure and 
energy density perturbations on comoving slices. 
Only for a universe dominated by a perfect fluid, we have $c_w = c_s$.

Note that the equation of motion for the Newtonian potential is 
well-defined even for the case in which both $w$ and $c_s$ vanish. 
In order to solve the above equation we need two matching conditions 
for $\Phi$ and $\Phi'$. As before, the continuity of the intrinsic and 
extrinsic curvatures at the surface of phase transition implies that 
$\Phi$ is continuous as given in Eq.~(\ref{Phi-bc}). 
On the other hand, using Eq.~\eqref{RPhi} and noting that both ${\calR}$
 and $\Phi$ are continuous across the transition one obtains the
 second matching condition,
\ba
\label{Phi-prime-bc}
\left[ \dfrac{\Phi'}{(1+w) \calH} + \dfrac{\Phi}{(1+w)}  \right]_\pm = 0  \, .
\ea 

Now we have the necessary information to obtain the solution for $\Phi$. 
At the first stage, the solution is 
\ba
\Phi_1 = A_1 \, x_1^{\nu_1-1} \, H^{(1)}_{\nu_1-1} (x_1) \, .
\ea
with the same definition of $x_1 (\eta) $ and $\nu_1$ as it is 
in Eq.~\eqref{xdef}. Using the relations~\eqref{PhiR} and \eqref{R1} 
one has
\ba
A_1 = - \dfrac{3}{2} \beta_1 (1+w_1) C_1  \, .
\label{coeffA1}
\ea
For the second stage with $w_2= c_{s2}=0$ , one has
\ba
\Phi_2 = A_2 \left[  \dfrac{\calH_{12}}{2} (\eta - \eta_{12})+ 1 \right]^{-5} 
+ B_2  \, .
\ea
Finally, for the third stage one has
\ba
\Phi_3 = x_3^{\nu_3-1} \left[  A_3 \, H^{(1)}_{\nu_3-1} (x_3)
   + B_3 \, H^{(2)}_{\nu_3-1} (x_3)   \right] \, .
\ea
Since we are interested in the power spectrum of the comoving
curvature perturbation at the final stage, we use Eq.~\eqref{RPhi}
to obtain 
\ba
\label{R-matter}
\calR_k = -\dfrac{2 \, x_3^{\nu_3}}{3 (1+w_3) \beta_3} 
\left[ A_3 \, H^{(1)}_{\nu_3} (x_3)+ B_3 \, H^{(2)}_{\nu_3} (x_3)\right] \,.
\ea
From this the transfer function is obtained as
\ba
\label{T-matter}
T^{1/2} = 
\dfrac{2 \vert A_3 - B_3 \vert}{3 |\beta_3|(1+w_3) \vert C_1 \vert}  \, .
\ea
The explicit expressions of $A_3$ and $B_3$ in terms of $C_1$ are 
computed in Appendix~{\ref{transfer-app}}. 

%%%%%%%%%%%%%%%%%%%%%%%%%%%%%%%%%%%%%%%%%%%%%%%%%%
\begin{figure}
\includegraphics[width =  4.5in ]{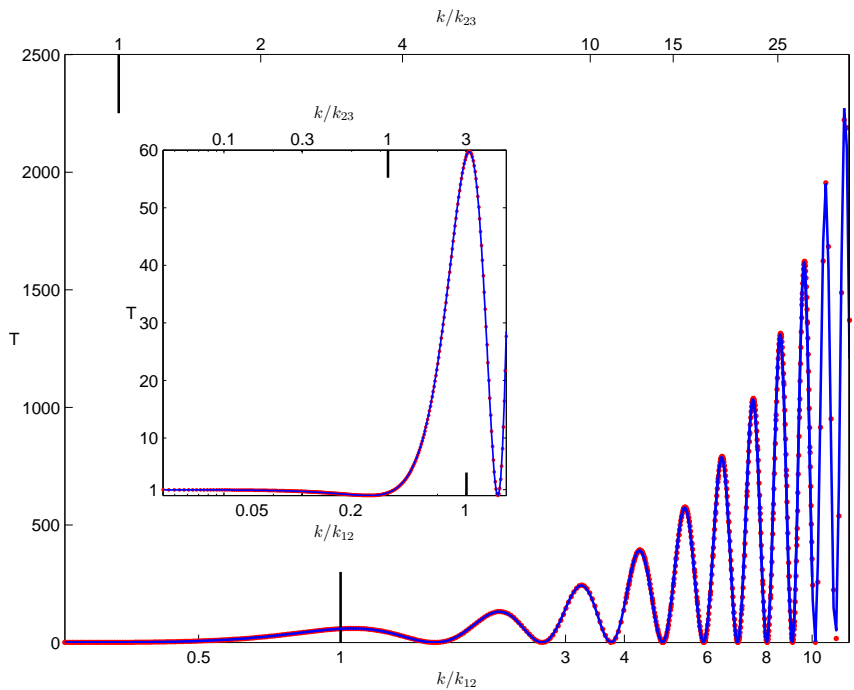}
\caption{The same plot as in \ref{T-radiation} but with $w_2=0$.   
\label{T-matter-fig}
}
\end{figure}
%%%%%%%%%%%%%%%%%%%%%%%%%%%%%%%%%%%%%%%%%%%%%%%%%%

In order to obtain an approximate expression for the transfer function, 
firstly note that $ \left({\calH_{23}}/{\calH_{12}}\right)^5 \ll 1 $ 
for $w_2\geq0$. Besides that all of $\beta_i$ and $w_i$ are of 
the order of unity. Using this information as well as the asymptotic
behavior of the Hankel functions, we find for the first category, $k>k_{12}$,
\ba
\label{T-matter1}
T^{1/2} \simeq \left( \dfrac{k}{k_{23}} \right)^{-\nu_3 + 3/2 }
 \left( \dfrac{k}{k_{12}} \right)^{\nu_1 - 1/2 } 
\dfrac{\sqrt{2} }{5 \beta_3 (1+w_3)} 
\sqrt{1+\sin\left(  \dfrac{2k}{k_{23}}  - \pi \nu_3\right)}\,;
\quad k> k_{12}\,.
\ea
and for the second category, $k_{23} < k < k_{12}$,
\ba
\label{T-matter2}
T^{1/2} \simeq \dfrac{2^{\nu_1 -2}}{5\sqrt{\pi}} 
\dfrac{\beta_1 (5+3 w_1)}{ \beta_3 (1+w_3)}\Gamma(\nu_1-1)  
\left( \dfrac{k}{k_{23}} \right)^{-\nu_3 + 3/2 } 
\sqrt{1+\sin \left( \dfrac{2 k}{k_{23}}  - \pi \nu_3 \right)} \,;
\quad k_{23} < k < k_{12}\,.
\ea

In Fig.~\ref{T-matter-fig} both numerical and analytic results 
are shown, which are in very good agreement with each other. 
Interestingly, the power spectrum on small scales is highly enhanced 
due to the sharp phase transition to the matter-dominated era. 
This agrees with the result obtained in \cite{Zaballa:2009xb},
and will have interesting implications for the primordial black hole 
formation~\cite{Carr:2009jm, Drees:2011hb, Drees:2011yz, Josan:2009qn}.

%%%%%%%%%%%%%%%%%%%%%%%%%%%%%%%%%%%%%%%%%%%%%%%%%%

\section{Discussions and Conclusions}
\label{conclusion}

In this work we considered a universe in which inflation has multiple 
stages, separated by intermediate non-inflationary stages. 
To be specific, we considered two inflationary stages separated
by an intermediate non-inflationary universe. To simplify the
calculation, we also assumed sharp transitions between the stages, 
and studied the case when the intermediate stage is either radiation 
dominated, matter dominated or kinetic energy dominated. 

To read off the final outgoing curvature perturbations, one has to perform the appropriate matching conditions at the surface of phase transitions. We have provided a consistent mechanism as how to do these matching conditions, given in Eqs. (\ref{R-bc1}) and (\ref{matchR'}) when $w \neq 0$ and  in Eq. (\ref{Phi-prime-bc}) when $w=0$. Furthermore, one has to take into account a non-trivial interplay between $c_s$ and $w$ when imposing the matching conditions which is captured by the parameter $f_{ij}$ as given in 
Eq. (\ref{Phi-prime-bc}) and the fact that $c_s \neq c_w$ as given in Eq. (\ref{Phi-matter}).
Some of these technical aspects of performing the matching conditions properly and the effects of $c_s$ have not been addressed in the previous works conclusively.

As well known, in the absence of isocurvature or entropy perturbations,
the comoving curvature perturbations are frozen on superhorizon scales,
$\dot{\cal R}=0$. So the transitions do not affect superhorizon modes.
However, sub-horizon modes are significantly affected.
Hence the final amplitude of the comoving curvature perturbation
depends crucially on the equation of state parameter 
of the intermediate stage, $w=w_2$,
as well as on the time when the mode of interest leaves the horizon.

%%%%%%%%%%%%%%%%%%%%%%%%%%%%%%%%%%%%%%%%%%%%%%%%%%
\begin{figure}
\includegraphics[scale=.45]{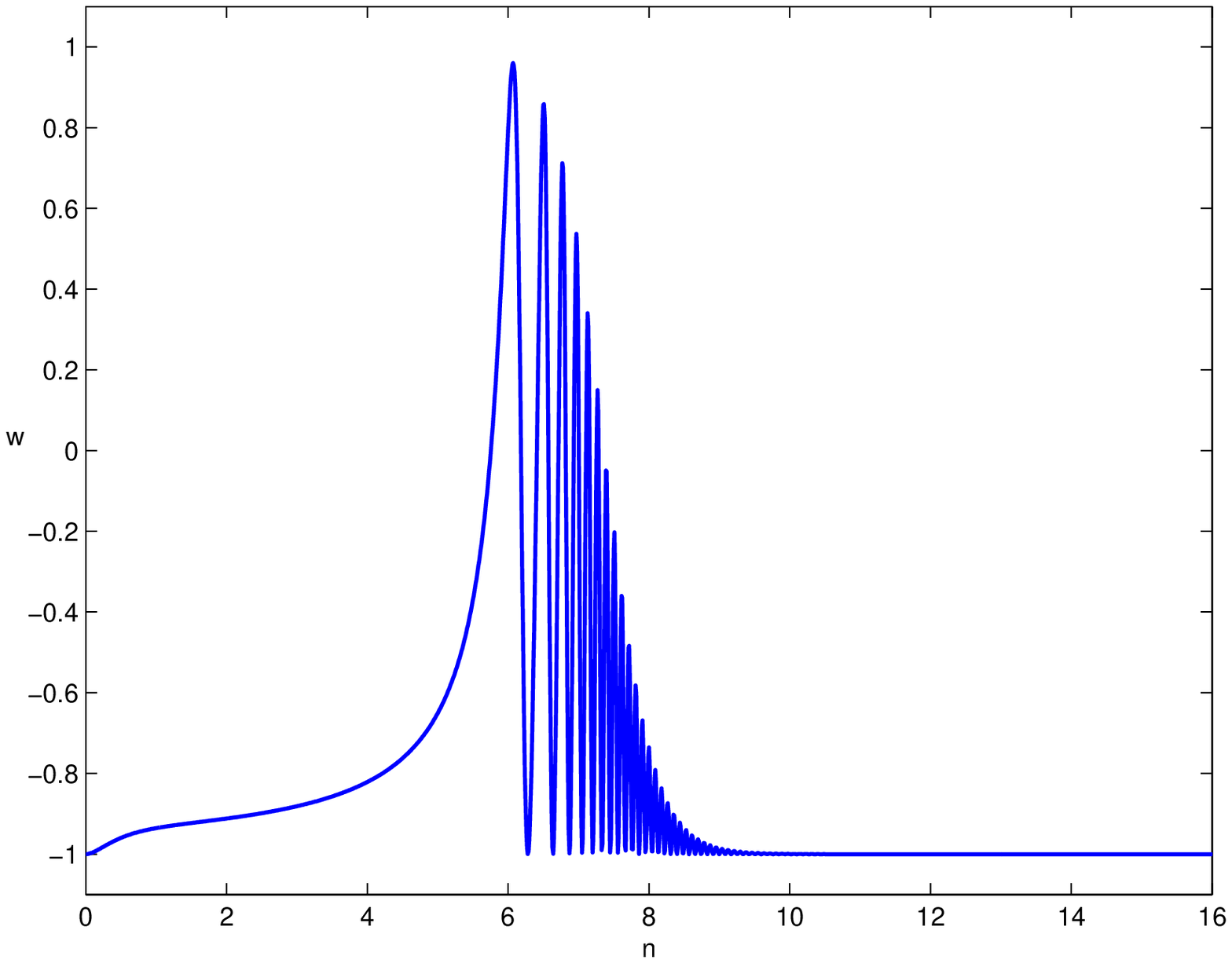}
\includegraphics[scale=.45]{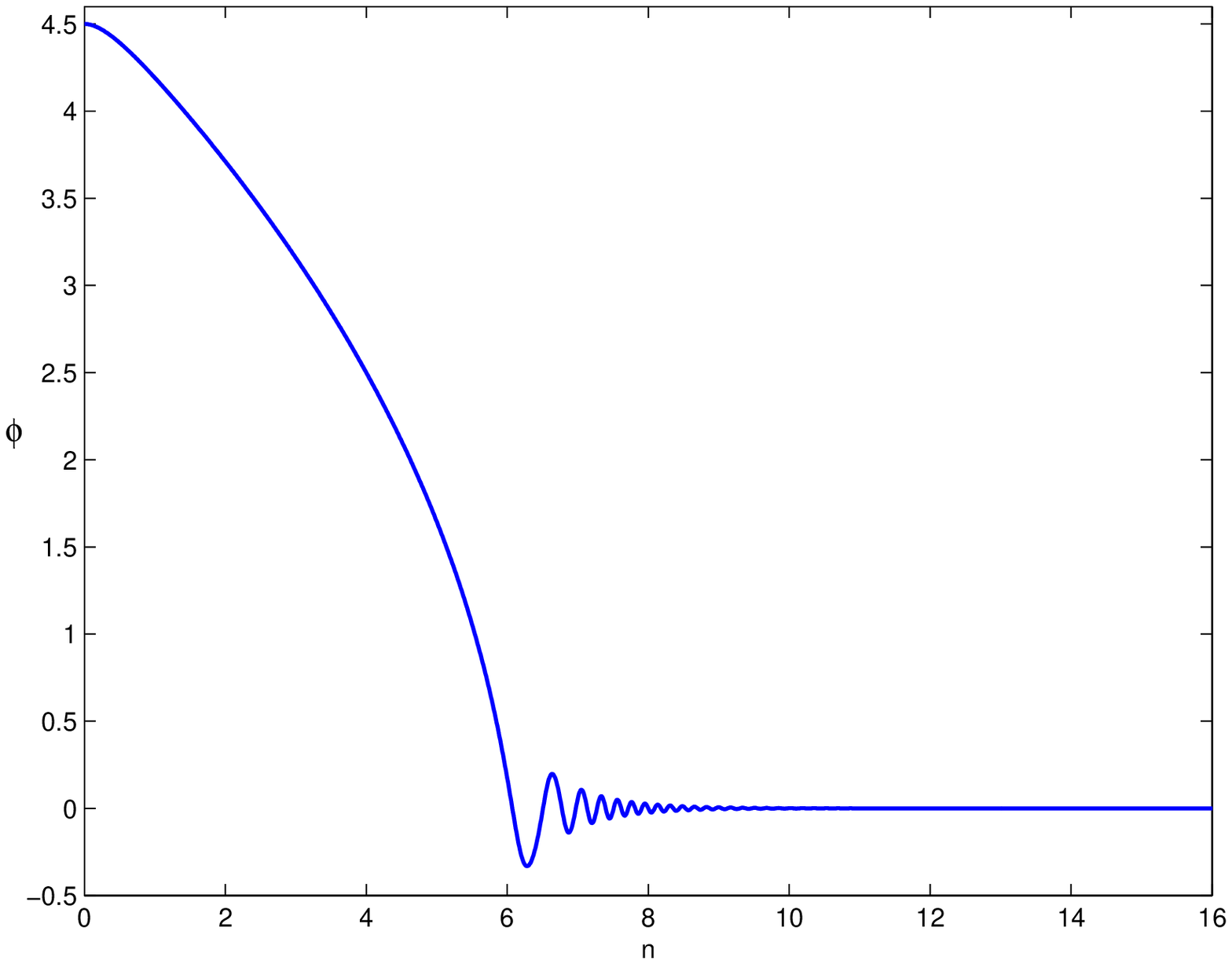}
\caption{The numerical plots for equation of state (Left) and 
the behavior of scalar field (Right) for the model in \eqref{V}. } 
\label{model-a}
\end{figure}  
%%%%%%%%%%%%%%%%%%%%%%%%%%%%%%%%%%%%%%%%%%%%%%%%%%

For modes which leave the horizon at the second stage of inflation, 
we see a sinusoidal modulations. If the phase transition
is arbitrarily sharp, then these sinusoidal modulations persist down
to infinitely small scales. This is an artifact of our assumption that
the change in $w$ is instantaneous. 
In a realistic situation in which the change in $w$ takes some finite 
time scale, then on sufficiently small scales the power spectrum retains
it pure vacuum form with no sinusoidal modulations. 

We found that the maximum enhancement in the power spectrum occurs 
for modes which leave the horizon either at the time of the
first phase transition $k=k_{23}$ or at the time of the second 
phase transition  $k= k_{12}$.
 For the case of a matter-dominated intermediate stage, $w_2=0$,
we found that the power spectrum increases towards small scales. 
This can have interesting implications for the primordial black-hole formation. 
It would be very interesting to study the possibility of the
over-production of primordial black-holes in our model.

In our analysis so far we have not provided a dynamical mechanism which 
causes a rapid change in the equations of state. Here it may be
worth mentioning a simple toy model which can mimic this situation.
 Consider a single scalar field with the canonical kinetic term 
which is slowly rolling down the potential. This corresponds
to the first inflationary stage.
When the inflaton field reaches near the minimum of the potential 
and starts to oscillate, the intermediate non-inflationary stage 
commences. The equation of state of this stage depends on the shape 
of the potential near its minimum. For a simple quadratic potential,
the universe behaves as matter dominated.

 Now we add a small cosmological constant to the potential
 which would not affect the dynamics of the first inflationary stage. 
After the energy density of the scalar field is diluted sufficiently 
due to damped oscillations, the cosmological constant starts to dominate
and the second stage of inflation begins.

Note that this is a single field scenario, so our previous analysis 
are applicable here.  However in order to terminate inflation, 
one may need an auxiliary field. Nevertheless,
one can assume that this field is sufficiently heavy so that 
it does not affect the large scale (CMB scale)  perturbations,
as in the hybrid inflation scenario~\cite{Linde:1993cn, Copeland:1994vg}.

\begin{figure}
\includegraphics[scale=.7]{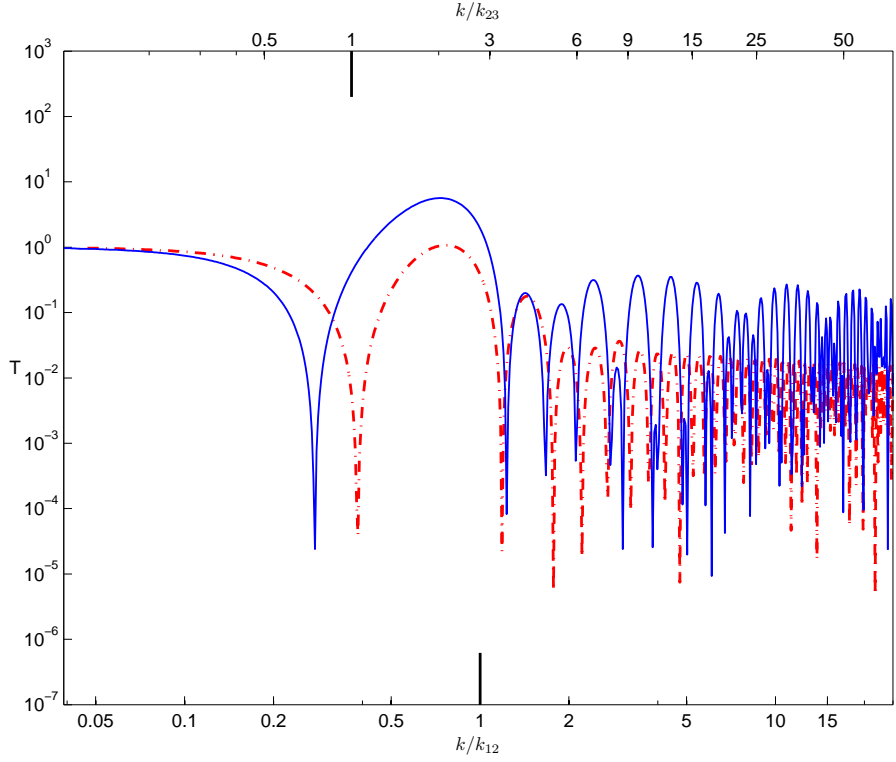}
\caption{Two numeric result for two different phase transitions. 
The solid blue line corresponds to a sharp phase transition while for 
the dashed red line the second transition is mild. In both cases we 
set $c_{s1}=c_{s2}=c_{s3}=1$. The other parameters are the same as
 in~\eqref{T-radiation}}
\label{model}
\end{figure}

The above model can be realized by the potential as simple as 
\ba 
\label{V}
V= \dfrac{1}{2} m^2 \phi^2 + V_0\,. 
\ea
We have used this for the numerical plots in Figs.~\ref{model-a} and \ref{model}.
As it is clear from the figure, the first phase transition can
be quite sharp, while the second transition turns out to be
relatively mild.
 Note that this case is different from what we have plotted in
 Fig.~\ref{T-matter-fig}, not only because of the mild second phase 
transition but also because the sound speed is equal to unity 
throughout all the stages, even for the intermediate stage.
This is because for the scalar field
 with the canonical kinetic term the sound speed is equal to unity.

%%%%%%%%%%%%%%%%%%%%%%%%%%%%%%%%%%%%%%%%%%%%%%%%%%
%%%%%%%%%%%%%%%%%%%%%%%%%%%%%%%%%%%%%%%%%%%%%%%%%%

If we relax the assumptions we adopted in the present paper,
and consider a multi-field or multi-fluid system,
then we may be able to construct more realistic models.
However the price to pay is the complication of the equations
of motion and the possible appearance of significant entropy/isocurvature
perturbations which are severely constrained by the current
observational data. Nevertheless it would be interesting to look for
such models.

%%%%%%%%%%%%%%%%%%%%%%%%%%%%%%%%%%%%%%%%%%%%%%%%%%
\section*{Acknowledgement}
We would like to thank A. A. Abolhasani,  N. Khosravi  and A. Mazumdar
for useful discussions.
This work was supported in part by MEXT Grant-in-Aid for 
the global COE program at Kyoto University,
"The Next Generation of Physics, Spun from Universality and Emergence,"
and by JSPS Grant-in-Aid for Scientific Research (A) No.~21244033.
This work was initiated while MHN was visiting the Yukawa Institute
for Theoretical Physics (YITP), Kyoto University, under the Exchange 
Program for Young Researchers of YITP.

%%%%%%%%%%%%%%%%%%%%%%%%%%%%%%%%%%%%%%%%%%%%%%%%%%%% July 9
\appendix

\section{Relations between $\Phi$ and ${\cal R}$}

\label{perts}

In this appendix we obtain the relations between $\Phi$ and ${\cal R}$ 
as given in Eqs. (\ref{RPhi}) and (\ref{PhiR}). 
The scalar perturbation of the metric is
\begin{equation}
ds^2 
= -(1+2A)dt^2 + 2a\partial_i B dx^idt + a^2[(1 + 2 F)\delta_{ij}
 +2\partial_i\partial_jE]dx^idx^j,
\end{equation}
The $00, ii, 0i$ and $i \neq j$ components of Einstein equations are
\ba
\label{00-Ein}
3 H  (\dot F - HA)  + \frac{k^2}{a^2}
\left[   F -  H \theta
\right]  &&= \frac{\delta \rho}{2 M_P^2}  \\
 \label{ii-Ein}
 ( 2 \dot H + 3 H^2)  A
+   H  \dot A - \ddot F -  3 H \dot F 
&& = \frac{\delta P}{2 M_P^2} \\
 \label{0i-Ein}
 -\dot F + H  A &&= -\frac{\rho +p}{2 M_P^2}  v\\
 \label{ij-Ein}
 \dot \theta + H  \theta -  A -F &&=0 
\ea
in which $\delta \rho$ and $\delta P$ are the energy density and 
the pressure perturbations. We have defined
\ba
\theta \equiv a^2 (\dot E - B/a) \, ,
\ea
and $v$ is the fluid velocity scalar potential defined as
$u_i = \partial_i v$ in which $u^\mu$ is the fluid four-velocity vector.  
A hypersurface on which $v=0$ is called the comoving slice.

The  gauge invariant variables Newtonian potential $\Psi$ and 
the Bardeen Potential $\Phi$ are given by 
\ba
\label{Phi-Psi}
\Psi = A - \dot \theta\,, \quad \Phi = F  - H \theta\,.
\ea
Furthermore, the curvature perturbation on constant energy hypersurface
 $\zeta$ and the curvature perturbation on comoving hypersurface ${\cal R}$
are given by
\ba
\zeta = F - \frac{H}{\dot{\rho}}\delta\rho  \,\quad 
{\cal R} = F + H v  \, .
\ea
We work in an isotropic background with no anisotropic pressure perturbation,
so from Eq.~(\ref{ij-Ein}) one concludes that $\Phi=- \Psi$.

To find Eqs.~(\ref{RPhi}) and (\ref{PhiR})  it is very helpful to go 
to the comoving slice on which $v=0$ and $F={\cal R}$.  In this gauge 
from Eq.~(\ref{0i-Ein}) we obtain $A= \dot{ \cal R}/H$. Plugging this into 
Eq.~(\ref{Phi-Psi}) to remove $\theta$ and using $\Phi= -\Psi$ yields
\ba
\label{RPhi2}
{\cal R} = \Phi -\frac{H}{\dot H} \left( \dot \Phi + H \Phi  \right) \, .
\ea
Using  $\dot H = -3 (1+ w) H^2/2$ and going to the conformal time one 
can easily check that Eq.~(\ref{RPhi2}) is equivalent to Eq.~(\ref{RPhi}).
To obtain Eq.~(\ref{PhiR}) we have to use the remaining two 
Einstein equations, Eqs.~(\ref{00-Ein}) and (\ref{ii-Ein}). 
The key point is that the sound speed of perturbations $c_s$ is defined on
the comoving slice as $\delta P_c = c_s^2 \delta \rho_c$ where
$\delta P_c$ and $\delta \rho_c$ are the pressure and energy density 
perturbations on the comoving slice.
Using this relation in Eqs.~(\ref{00-Ein}) and (\ref{ii-Ein}) and 
noting that $A= \dot{ \cal R}/H$ on this slice, one obtains
\ba
\label{PhiR2}
\Phi  = \frac{a^2 \dot H}{H k^2  c_s^2 }  \dot {\cal R} \, .
\ea
As above, using $\dot H = -3 (1+ w) H^2/2$ and going to the conformal 
time one can easily check the equivalence between  Eq.~(\ref{PhiR2}) 
and Eq.~(\ref{PhiR}).

%%%%%%%%%%%%%%%%%%%%%%%%%%%%%%%

%%%%%%%%%%%%%%%%%%%%%%%%%%%%%%%%%%%%%%%%%%%%%%%%%%%%%%%
\section{Asymptotic behavior of the Hankel function}
\label{apx:Hankel}

Here we recapitulate the asymptotic behavior of the Hankel functions. 
For large values of the argument one has
\ba
\label{large-Hankel}
H^{(1)}_\nu(Z \gg 1) &\simeq & \sqrt{\dfrac{2 }{i \pi Z}}\,
e^{i(Z-\pi \nu/2)} \, ,
\ea
whereas for small values of the argument,
\ba
\label{small-Hankel}
H^{(1)}_\nu(Z\ll 1) &\simeq & -\Gamma(\vert \nu \vert) 
\left( \dfrac{i}{\pi} \right)  \left( \dfrac{2}{Z} \right)^{\vert \nu \vert}
 e^{-i \frac{\pi}{2} (\nu-\vert \nu \vert)} \, .
\ea
Also the following identities are helpful:
\ba
\label{H-prime}
H'_\nu(Z) = H_{\nu-1}(Z) - \dfrac{\nu}{Z} H_\nu(Z)  \, ,
\ea
and
\ba
\label{Wronskian}
{H_\nu^{(1)}}(Z) {H_\nu^{(2)}}'(Z)
 -  {H_\nu^{(1)}}'(Z) {H_\nu^{(2)}}(Z) = -\frac{4 i}{\pi Z}\, .
\ea

%%%%%%%%%%%%%%%%%%%%%%%%%%%%%%%%%%%%%%%%%%%%%%%%%%
\section{Transfer Function for Outgoing Solutions}
\label{transfer-app}

By solving the matching conditions here we calculate the coefficients of outgoing solutions $C_2, D_2, C_3$ and $D_3$ for $w_2 \neq 0$ and $A_2, B_2, A_3$ and $B_3$ for $w_2$=0 
in terms of $C_1$. We consider each case separately.

%%%%%%%%%%%%%%%%%%%%%%%%%%%%%%%%%%%%%%%%%%%%%%%%%%

\subsection{$w_2 \neq 0$}

Consider the case where the intermediate non-inflationary stage is not 
matter dominated, $w_2=c_{s2}^2\neq 0$. The general solution for 
the comoving curvature perturbation at the intermediate non-inflationary
and second inflationary stages are given in Eqs.~(\ref{R2}) and (\ref{R3}),
respectively. 
With the help of the identities~(\ref{H-prime}) and (\ref{Wronskian}) 
and imposing the matching conditions~(\ref{R-bc1}) and (\ref{matchR'}) 
at $\eta_{12}$ and $\eta_{23}$, we obtain 
\ba
\label{exact1}
C_2&=& -\dfrac{\pi \, x_1(\eta_{12})^{\nu_1}}{4 i \, x_2(\eta_{12})^{\nu_2-1}}
C_1 \left[H^{(1)}_{\nu_1} \left(x_1(\eta_{12}) \right) 
H^{(2)}_{\nu_2-1}\left(x_2(\eta_{12})\right)
- f_{12} H^{(1)}_{\nu_1-1} (x_1(\eta_{12})) 
H^{(2)}_{\nu_2} \left(x_2(\eta_{12})\right) \right]\,,
\\ 
\label{exact2}
D_2 &=&\dfrac{\pi \, x_1(\eta_{12})^{\nu_1}}{4 i \, x_2(\eta_{12})^{\nu_2-1}}
 C_1 \left[H^{(1)}_{\nu_1} (x_1(\eta_{12})) 
H^{(1)}_{\nu_2-1}(x_2(\eta_{12}))- f_{12} 
H^{(1)}_{\nu_1-1} (x_1(\eta_{12})) H^{(1)}_{\nu_2}
 (x_2(\eta_{12})) \right]\,,
\\
\nonumber\\
\label{exact3}
C_3 &=& -\dfrac{\pi \, x_2(\eta_{23})^{\nu_2}}{4 i \, x_3(\eta_{23})^{\nu_3-1}}
\left[ C_2 \left(  H^{(1)}_{\nu_2} (x_2(\eta_{23})) 
H^{(2)}_{\nu_3-1}(x_3(\eta_{23})) - f_{23}  
H^{(1)}_{\nu_2-1} (x_2(\eta_{23})) 
H^{(2)}_{\nu_3} (x_3(\eta_{23})) \right) \right.  
\nonumber\\
&&\qquad\qquad\qquad 
+\left. D_2 \left(  H^{(2)}_{\nu_2} (x_2(\eta_{23})) 
H^{(2)}_{\nu_3-1}(x_3(\eta_{23}))- f_{23}  
H^{(2)}_{\nu_2-1} (x_2(\eta_{23})) 
H^{(2)}_{\nu_3}(x_3(\eta_{23})  \right) \right]\,,
\\
\nonumber\\
\label{exact4}
D_3 &=& \dfrac{\pi \, x_2(\eta_{23})^{\nu_2}}{4 i \, x_3(\eta_{23})^{\nu_3-1}}
\left[ C_2 \left(  H^{(1)}_{\nu_2} (x_2(\eta_{23})) 
H^{(1)}_{\nu_3-1}(x_3(\eta_{23})) - f_{23}  
H^{(1)}_{\nu_2-1} (x_2(\eta_{23})) 
H^{(1)}_{\nu_3} (x_3(\eta_{23})) \right) \right.  
\nonumber\\
&&\qquad\qquad\qquad
 +\left. D_2 \left(  H^{(2)}_{\nu_2} (x_2(\eta_{23}))
H^{(1)}_{\nu_3-1}(x_3(\eta_{23}))- f_{23}  
H^{(2)}_{\nu_2-1} (x_2(\eta_{23})) 
H^{(1)}_{\nu_3}(x_3(\eta_{23})  \right) \right] \,,
\ea
where we have defined 
\ba
f_{ij}=  \frac{\mathrm{sgn}(1+ 3 w_i)}{\mathrm{sgn}(1+ 3w_j)}
 \dfrac{(1+w_i)}{(1+w_j)}  
\frac{c_{sj}}{c_{si}} \, .
\ea
Note that in the continuous limit where  $w_1 \rightarrow w_2$ so 
$f_{12}\rightarrow  1$, we obtain the expected results that 
$C_2=C_1$ and $D_2 =0$. It is interesting to note that
it is the combination $f_{ij}$ of the changes in $w$ and $c_s$
that controls whether or not we have a non-trivial phase transition.

We would like to calculate the power spectrum of the modes which 
are super-horizon at the end of inflation, $\eta_e =0$, corresponding
 to $x_3(\eta_e) \ll 1$.  The transfer function is defined in 
Eq.~(\ref{T-def}). As described in the text we divide the modes of 
interest into three categories. The first category is defined by
$k> k_{12}$, the second category by $k_{23} <k< k_{12}$
and the third category by $k < k_{23}$. 

First consider the first category, $k > k_{12}$. For these modes
we have $ x_1(\eta_{12})$, $x_2(\eta_{12})$, $x_2(\eta_{23})$, 
$x_3(\eta_{23})\gg 1$. 
Using the large value approximation of the Hankel functions we obtain
\ba
\label{approx1}
C_2 &&\simeq  i D_2 \left(\dfrac{f_{12}+1}{-f_{12}+1} \right) 
e^{-2i (x_2(\eta_{12})-\pi \nu_2/2)}
\simeq
 \dfrac{C_1 x_1(\eta_{12})^{\nu_1-1/2}}{2 \, x_2(\eta_{12})^{\nu_2-1/2}}
(f_{12}+1)  e^{i(x_1(\eta_{12})-x_2(\eta_{12})-\pi (\nu_1-\nu_2)/2)}\,,
\\ 
C_3 &&\simeq  i D_3  \dfrac{{\cal I}}{{\cal J}}\, 
 e^{-2i (x_3(\eta_{23})-\pi \nu_3/2)} 
\\ \nonumber
&&\simeq {\cal I}
\dfrac{C_2 \, x_2(\eta_{23})^{\nu_2-1/2}}{2 \, x_3(\eta_{23})^{\nu_3-1/2}} 
 \, e^{i(x_2(\eta_{23})-x_3(\eta_{23})-\pi (\nu_2-\nu_3)/2)}  \, ,
\ea
where
\ba
{\cal I} &\equiv&
(1+f_{23}) \left[1+\dfrac{1-f_{23}}{1+f_{23}} \, \dfrac{1-f_{12}}{1+f_{12}} 
\, e^{2 i (x_2(\eta_{12})-x_2(\eta_{23}))} \right] 
\simeq f_{23} \left[1- e^{2 i (x_2(\eta_{12})-x_2(\eta_{23}))}\right]\,,
\nonumber
\\
{\cal J}& \equiv&
 (1-f_{23}) \left[1+\dfrac{1+f_{23}}{1-f_{23}} \, \dfrac{1-f_{12}}{1+f_{12}} 
\, e^{2 i (x_2(\eta_{12})-x_2(\eta_{23}))} \right] 
\simeq -{\cal I} \,.
\ea
The last approximate equality is valid for the case in which 
$w_1 \sim w_3 \simeq  -1$ and as a result $f_{12}<<1$ and $f_{23}>>1$.
 Plugging these values of $C_3$ and $D_3$ in 
the formulas for the final curvature perturbation amplitude~(\ref{R-end}) 
and the transfer function~(\ref{T-def}), we obtain 
Eqs.~(\ref{R-cat1}) and (\ref{T-cat1}).

Now consider the second category, $k_{23} < k < k_{12}$.
For these modes we have $x_1(\eta_{12})$, $x_2(\eta_{12}) \ll 1 $ 
while $x_2(\eta_{23})$, $x_3(\eta_{23}) \gg 1$. 
For this category, we obtain
\ba
\label{approx2}
C_2 &\simeq& D_2 \, e^{2 i \pi \nu_2}  \simeq 
\dfrac{-i}{\pi} 2^{\nu_1-\nu_2-1} C_1 \Gamma(\nu_1) 
\Gamma(-\nu_2+1) e^{-i \pi \nu_2}\,,
\\ 
C_3 &\simeq& i D_3 \dfrac{{\cal I}'}{{\cal J}'} e^{-2 i ( x_3-\pi \nu_3/2)} 
\simeq 
 \dfrac{{\cal I}'}{2} C_2 \, x_3(\eta_{23})^{-\nu_3+1/2} 
x_2(\eta_{23})^{\nu_2-1/2} 
 e^{i(x_2(\eta_{23})-x_3(\eta_{23})-\pi (\nu_2-\nu_3)/2)}\,,
\ea
where
\ba
 {\cal I}' &=& ( f_{23}+1)+i (1-f_{23}) e^{-2 i (x_2(\eta_{23})-3 \pi \nu_2/2)}
\simeq f_{23} \left(1-i e^{-2 i (x_2(\eta_{23})-3 \pi \nu_2/2)} \right) \,,
\\ \nonumber
{\cal J}'& =& (-f_{23}+1)+i (1+f_{23}) e^{-2 i (x_2(\eta_{23})-3 \pi \nu_2/2)}
\simeq -{\cal I}'\,.
\ea
Again the second approximate equality holds for the case $f_{23} \gg 1$. 
Now plugging these expressions of $C_3$ and $D_3$ into the formulas
for the final curvature perturbation amplitude~(\ref{R-end}) and 
the transfer function~(\ref{T-def}), we obtain Eqs.~(\ref{Rsecond}) and
(\ref{Tsecond}).

%%%%%%%%%%%%%%%%%%%%%%%%%%%%%%%%%%%%%%%%%%%%%%%%%%
\subsection{Matter Dominated Intermediate Era}

Now consider the case where the intermediate non-inflationary stage is a 
matter-dominated universe, so $w_2 = c_{s2}=0$. 
As described in the text for this case we work with the 
Bardeen potential $\Phi$ while for the final curvature perturbation 
power spectrum we can switch to ${\cal R}$ as given by Eq.~(\ref{R-matter})
by using Eq.~(\ref{RPhi}).
Our task is to find the expressions of $A_3$ and $B_3$ in Eq.~(\ref{R-matter})
in terms of $A_1$, which is expressed in terms of $C_1$ in Eq.~(\ref{coeffA1}).

Imposing  the matching conditions (\ref{Phi-bc}) and (\ref{Phi-prime-bc}) 
at $\eta = \eta_{12}$ and $\eta = \eta_{23}$ yields
\ba
A_2 &=& 
- \dfrac{2 w_1 \,  A_1 \, x_1(\eta_{12})^{\nu_1-1}}{5 (1+w_1)} 
\left[ \dfrac{x_1(\eta_{12})}{\beta_1 w_1} 
H^{(1)}_{\nu_1 -2}(x_1(\eta_{12})) - H^{(1)}_{\nu_1-1}(x_1(\eta_{12}))\right]\,,
\\  
B_2 &=&
 A_1 x_1(\eta_{12})^{\nu_1-1}  H^{(1)}_{\nu_1 -1}(x_1(\eta_{12}))  -A_2\,,
 \\
A_3 &=& 
- \dfrac{\pi}{8 i}  x_3(\eta_{23})^{1-\nu_3} \left[ 2  x_3(\eta_{23}) 
\left(  B_2  + A_2 \left(\dfrac{\calH_{23}}{\calH_{12}} \right)^5\right) 
\, H^{(2)}_{\nu_3-2} (x_3(\eta_{23}))  \right.
\nonumber\\
 &&+ \left. \beta_3 \left( -2 w_3 B_2 + (5+3 w_3) A_2 
\left(\dfrac{\calH_{23}}{\calH_{12}} \right)^5 \right) 
H^{(2)}_{\nu_3-1} (x_3(\eta_{23})) \right]\,,
\\
B_3 &=&
\dfrac{\pi}{8 i}  x_3(\eta_{23})^{1-\nu_3} \left[ 2  x_3(\eta_{23}) 
\left(  B_2  + A_2 \left(\dfrac{\calH_{23}}{\calH_{12}} \right)^5\right) 
\, H^{(1)}_{\nu_3-2} (x_3(\eta_{23})) 
\right.
\nonumber\\
&&+\left.  \beta_3 \left( -2 w_3 B_2 + (5+3 w_3) A_2 
\left(\dfrac{\calH_{23}}{\calH_{12}} \right)^5 \right) 
H^{(1)}_{\nu_3-1} (x_3(\eta_{23}))  
\right]\,.
\ea

As before, we evaluate $A_3$ and $B_3$ for the three different categories
of the modes separately. 
For the first category, $k> k_{12}$, we have
\ba
A_2 &\simeq& B_2 \simeq 
- \dfrac{2i A_1 x_1(\eta_{12})^{\nu_1-1/2}}{5 \beta_1 (1+w_1)}
 \sqrt{\dfrac{2i}{\pi} } e^{i(x_1-\pi \nu_1/2 )}  \,,
\\
A_3 &\simeq& -i e^{-2i( x_3(\eta_{23}) - \pi \nu_3/2  )} B_3 \, 
\simeq \sqrt{\dfrac{2\pi i}{8}} \, B_2 \, x_3(\eta_{23})^{-\nu_3+3/2}
 e^{-i(x_3(\eta_{23})-\pi \nu_3/2)} \,.
\ea
For the second category, $k_{23} < k< k_{12}$, we have
\ba
A_2 &\simeq& \dfrac{2}{3+5 w_1} B_2 
\simeq \dfrac{-i 2^{\nu_1} w_1}{5 \pi (1+w_1)} A_1 \Gamma(\nu_1-1)\,,
\\
A_3 &\simeq& -i e^{-2i( x_3(\eta_{23}) - \pi \nu_3/2  )} B_3 \, 
\simeq \dfrac{\pi}{4} \, \sqrt{\dfrac{2i}{\pi}} \, B_2 \, 
x_3(\eta_{23})^{-\nu_3+3/2} e^{-i(x_3(\eta_{23})-\pi \nu_3/2)} \,.
\ea
Plugging these into the transfer function formula~(\ref{T-matter}) yields 
Eqs.~(\ref{T-matter1}) and (\ref{T-matter2}).

%%%%%%%%%%%%%%%%%%%%%%%%%%%%%%%%%%%%%%%%%%%%%%%%%%

\section{Necessary relations for numeric calculations}
In order to have an efficient and accurate code, it is much better to 
change the variable from the conformal time to the number of $e$-folds, 
\ba
dn=\calH d\eta \,.
\ea
Using the Friedmann equation one has
\ba
\calH(n)=\calH_e \, \exp\left( \dfrac{1}{2} 
\int^{n_e}_{n} (1+3 w) \, dn' \right)\,,
\ea
where $n_e-n$ is the total number of $e$-folds 
from the time $n$ until the end of inflation.

The curvature perturbation obeys the equation of motion in terms of $n$,
\ba
\dfrac{d^2 \calR}{dn^2}+\left(\dfrac{3}{2} (1-w)
+\dfrac{d}{dn}\ln \left((1+w)/c_s^2 \right) \right) 
\dfrac{d\calR}{dn}+ \dfrac{c_s^2 k^2}{\calH^2} \calR=0\,.
\ea
Similarly one has
\ba
\dfrac{d^2 \Phi }{dn^2}  + \left[\dfrac{5}{2} + 3 (c_w^2 - w/2 ) \right]
\dfrac{d \Phi }{dn}
 + \left[\dfrac{c_s^2 k^2}{\calH^2} - 3 (w- c_w^2) \right] \Phi =0\,.
\ea
To model smooth transitions, we use the error function. 
For example the equation of state is given by
\ba
w=w_1 + (-w_1 + w_2)\left(1 + {\rm{erf}}[d_1 (n - n_1)]\right)/2 
+ (-w_2 + w_3)\left(1 + {\rm{erf}}[d_2 (n - (n_1+n_2))]\right)/2\,,
\hspace{.5cm}
\ea
where $d_1$ and $d_2$ determine the sharpness of the first and 
second phase transitions, respectively.

%%%%%%%%%%%%%%%%%%%%%%%%%%%%%%%%%%%%%%%%%%%%%%%%%%
{}

\end{document}